\documentclass[graphics,floatfix,footinbib,nobibnotes, twocolumn]{revtex4-1}
\usepackage{amsmath,amssymb,wasysym}
\usepackage{graphicx}
\usepackage{dcolumn}
\usepackage{bm}
\usepackage{braket}
\usepackage{subcaption}
\usepackage{verbatim}
\usepackage{float}
\usepackage{bbold}
\usepackage{color}
\usepackage{xcolor}
\usepackage[colorlinks=true ,urlcolor=blue,urlbordercolor={0 1 1}]{hyperref}
\usepackage{cleveref}
\usepackage{relsize}
\usepackage{amsthm}
\usepackage{enumerate}
\usepackage{soul,xcolor}
\usepackage[T1]{fontenc}
\usepackage{adjustbox}

\renewcommand{\v}[1]{\mathbf{#1}} 

\newcommand{\be}{\begin{equation}}
\newcommand{\ba}{\begin{align}}
\newcommand{\ee}{\end{equation}}
\newcommand{\bea}{\begin{eqnarray}}
\newcommand{\eea}{\end{eqnarray}}
\newcommand{\beq}{\begin{equation}}
\newcommand{\eeq}{\end{equation}}
\newcommand{\beqn}{\begin{eqnarray}}
\newcommand{\eeqn}{\end{eqnarray}}

\renewcommand{\hat}[1]{{\widehat #1}}

\begin{document}
\title{Phase transitions out of quantum Hall states in moir\'e materials}
\author{Xue-Yang Song}
\affiliation{Department of Physics, Massachusetts Institute of Technology, Cambridge MA 02139,  USA}

\author{Ya-Hui Zhang}
\affiliation{Department of Physics and Astronomy, Johns Hopkins University, Baltimore, Maryland 21218, USA}

\author{T. Senthil}
\affiliation{Department of Physics, Massachusetts Institute of Technology, Cambridge MA 02139-4307,  USA}

\date{\today}

\begin{abstract}
Motivated by the recent experimental breakthroughs in observing Fractional Quantum Anomalous Hall (FQAH) states in moir\'e  materials, we propose and study various unconventional phase transitions between quantum Hall phases and  Fermi liquids or charge ordered phases upon tuning the bandwidth.  At a fixed rational lattice filling $\nu$, we describe a quantum Ginzburg-Landau theory to describe the intertwinement between the  FQAH and Charge Density Wave (CDW) orders.  We use this theory to describe phase transitions between the FQAH and a CDW insulator.   The critical theory for a direct second order transition resembles that of the familiar deconfined quantum critical point (DQCP) but with an additional Chern-Simons term. At filling $\nu=-\frac{1}{2}$, we study the possibility of a continuous transition between the composite Fermi liquid (CFL) and the Fermi liquid (FL) building on and refining previous work by  Barkeshli and McGreevy (\href{https://journals.aps.org/prb/abstract/10.1103/PhysRevB.86.075136}{Phys. Rev. B 86, 075136}).    Crucially  we show that filling constraints ignored in that work ensure that translation symmetry alone is enough to enable a second order CFL-FL transition. We argue that there must be  critical CDW fluctuations though neither phase has long range CDW order.  We present experimental signatures the most striking of which is a universal jump of both longitudinal and Hall resistivities at the critical point. With disorder, we argue that the CDW order gets pinned and the CFL-FL evolution happens through an intermediate electrically insulating phase with mobile neutral fermions. A clean analog of this insulating phase with long range CDW order and a neutral fermi surface can potentially also exist. We discuss the properties of this phase and the nature of its phase transitions.   We also present a critical theory for the CFL to FL transition  at filling $\nu=-\frac{3}{4}$.  Our work opens up a new avenue to realize deconfined criticality and fractionalized phases beyond  familiar Landau level physics in the moir\'e Chern band system.

\end{abstract}

\maketitle

\section{Introduction}
The phenomena associated with the quantum Hall regime are amongst the most fascinating in modern condensed matter physics. For electrons moving in a strong magnetic field in two dimensions, at fractional filling of the lowest Landau level, many correlated states have been observed. These include incompressible fractional quantum Hall (FQH) states\cite{stormer1999fractional,jain1989composite} and compressible metallic non-fermi liquid states known as composite fermi liquids (CFL) at half and quarter fillings\cite{halperin1993theory}.  Relatively recently it was pointed out that fractional quantum Hall physics may also be achieved at zero magnetic field in a narrow Chern band\cite{sun2011nearly,sheng2011fractional,neupert2011fractional,wang2011fractional,tang2011high,regnault2011fractional,bergholtz2013topological,parameswaran2013fractional}. In this avatar, the corresponding phases have been dubbed `Fractional Quantum Anomalous Hall' (FQAH) states. Two dimensional moir\'e systems were proposed to be a particularly suitable platform to host nearly flat Chern bands and thus to realize  FQAH states\cite{zhang2019nearly,ledwith2020fractional,repellin2020chern,abouelkomsan2020particle,wilhelm2021interplay}.  Indeed integer quantum anomalous Hall states\cite{sharpe2019emergent,serlin2020intrinsic} have been observed in twisted bilayer graphene(TBG) aligned with hexagon boron nitride (hBN)\cite{zhang2019twisted,bultinck2020mechanism}, in ABC trilayer graphene aligned with hBN\cite{chen2020tunable}, and in some moir\'e semiconducting Transition Metal Dichalcogenide (TMD) materials\cite{li2021quantum,foutty2023mapping}.   Fractional Chern insulators have been observed in some moir\'e platforms in non-zero magnetic fields\cite{spanton2018observation,xie2021fractional}. 

Very recently, a fractional quantum anomalous Hall effect (FQAH) was found in twisted MoTe$_2$ moir\'e heterostructures, first through compressibility measurements\cite{cai2023signatures,zeng2023integer}, and then directly in transport\cite{park2023observation}.  Theoretically it had previously been shown that twisted  TMD homobilayers may host a Chern band\cite{wu2019topological,yu2020giant,devakul2021magic} and FQAH states \cite{li2021spontaneous,crepel2023fci} through a spontaneous breakdown of time-reversal symmetry associated with polarization of the valley degree of freedom. After the experimental observation, several theoretical works confirmed the existence of FQAH from numerical studies\cite{wang2023fractional,reddy2023fractional}. Further the numerical results show that valley-polarized composite Fermi liquids (CFL) can  exist \cite{dong2023composite,goldman2023zero} at $1/2$ and $3/4$ fillings, in agreement with the simple picture that the flat Chern band in twisted MoTe$_2$ mimics the lowest Landau level quite well.

What new physics, beyond what is accessible in conventional FQH platforms, can we pursue in these moir\'e realizations? Perhaps the most interesting new capability afforded by the moir\'e setting is the tunability of parameters. First the band filling can be easily controlled by an external gate. More remarkably, through a perpendicular displacement field, the band width of the Chern band can be controlled while preserving the interaction strength, as already demonstrated experimentally\cite{cai2023signatures,zeng2023integer,park2023observation}. For correlated states in a Chern band, we can thus hope to study the evolution between quantum Hall physics and a variety of other phases by tuning the displacement field at fixed filling. Examples of the latter include conventional Landau Fermi liquid metals, and charge-ordered insulators. 

Here we present a general theory for the competition/intertwinement  between incompressible FQAH states, and CDW insulators. We derive a `quantum' Ginzburg-Landau theory for this competition that naturally incorporates both phases and enables discussing the phase transition. For the prominent incompressible FQAH state at $\nu = -2/3$, we show that a continuous quantum phase transition to a Charge Density Wave (CDW) insulator (with a coexisting Integer Quantum Anomalous Hall effect) is potentially possible and describe its physical properties. A different transition from the $\nu = -\frac{2}{3}$ FQAH state to a CDW insulator without a quantized electrical Hall conductivity is also possible; the CDW insulator has a coexisting `dark' topological order with electrically neutral anyons which will not affect electrical transport measurements.

At fillings $\nu=-1/2$ and $\nu = -3/4$,  apart from compressible composite fermi liquid states, two other prominent possible states are a Landau Fermi liquid (FL), and CDW insulating states that break lattice translation symmetry. We will discuss the associated phase transitions between these phases. 
A theory for a continuous CFL to FL transition was proposed previously by Barkeshli and McGreevy\cite{barkeshli2014continuous}. This theory is based on a parton construction where the electron operator $c$ is written as $c = \Phi f$ whhere $\Phi$ is a boson and $f$ is a fermion. $\Phi$ and $f$ are both coupled to a dynamical $U(1)$ gauge field with equal and opposite charges. As usual if $f$ forms a Fermi surface state, and $\Phi$ forms a superfluid, then the electrons form the conventional Landau fermi liquid. If however $\Phi$ undergoes a transition to the $\nu = 1/2$ Laughlin state, then the resulting electronic state is a composite fermi liquid described by the Halperin-Lee-Read effective theory. The phase transition between these two states is thus described as a phase transition of bosons between Laughlin and superfluid states in the presence of a coupling to a dynamical $U(1)$ gauge field which, in turn, is also coupled to the fermi surface of the $f$-fermions. The bosonic Laughlin-superfluid transition  is itself an interesting theoretical problem studied earlier by Barkeshli and McGreevy\cite{barkeshli2012continuous} who showed that it is plausibly second order (and described by a conformal field theory). The coupling of this theory to the gapless gauge field and $f$-fermi surface is then treated following methods used in the treatment of a continuous Mott transition between a Fermi liquid and spin liquid Mott insulator by one of us in Ref. \onlinecite{senthil2008theory}. 

  This prior work needs several refinements which we supply in the present paper. Specifically, Refs. \onlinecite{barkeshli2014continuous,barkeshli2012continuous} did not consider the restrictions coming from the lattice filling of $1/2$ an electron per site on average. Such filling constraints play a crucial role in determining the structure of the low energy effective field theory of correlated electrons. We will incorporate these filling constraints and show that they lead to some crucial new features absent in this prior work.   We also address a number of theoretical questions left open which we summarize below, and  present concrete suggestions for experiments that could probe the physics of the CFL-FL transition.  

 For the building block of the bosonic Laughlin-superfluid transition, Barkeshli and McGreevy\cite{barkeshli2012continuous} proposed a description of it as a theory of two massless Dirac fermions coupled to a $U(1)$ gauge field with a Chern-Simons term.  This theory  has  three relevant operators corresponding to bilinears of the Dirac fermions $\bar \psi \vec \sigma \psi$, which, if allowed,  will split the direct transition  and lead to an intermediate phase.  Therefore, in order to have a direct continuous transition, there needs to be a crystal symmetry to forbid these terms.  This issue was not  studied in a concrete setup in Ref.~\onlinecite{barkeshli2014continuous} and Ref.~\onlinecite{barkeshli2012continuous}.  Given the low symmetry associated with a single valley in the twisted MoTe$_2$ structure (just lattice translation and $C_3$ rotations), it is thus not a priori clear whether there can be a continuous Laughlin-superfluid transition for bosons (and the associated CFL-FL transition for fermions) at half filling. Here,  we show that lattice translation symmetry and associated filling constraints  (which were not incorporated in earlier work) are enough to prohibit these relevant operators, thereby making a second order transition more plausible. Indeed these operators describe fluctuating CDW order that rears its head just at the critical point though static CDW order is absent in either phase. These CDW orders live at the three {\bf M} points of the hexagonal Brillouin zone, and are transformed into each other by the microscopic $C_3$ symmetry present in a single valley. 
We argue that a weak breaking of  the $C_3$ symmetry is an irrelevant perturbation, and hence the continuous CFL-FL transition is stable even in the presence of weak strain.

\begin{figure}
\adjustbox{trim={.08\width} {.48\height} {.1\width} {.1\height},clip}
   {\includegraphics[width=.8\textwidth]{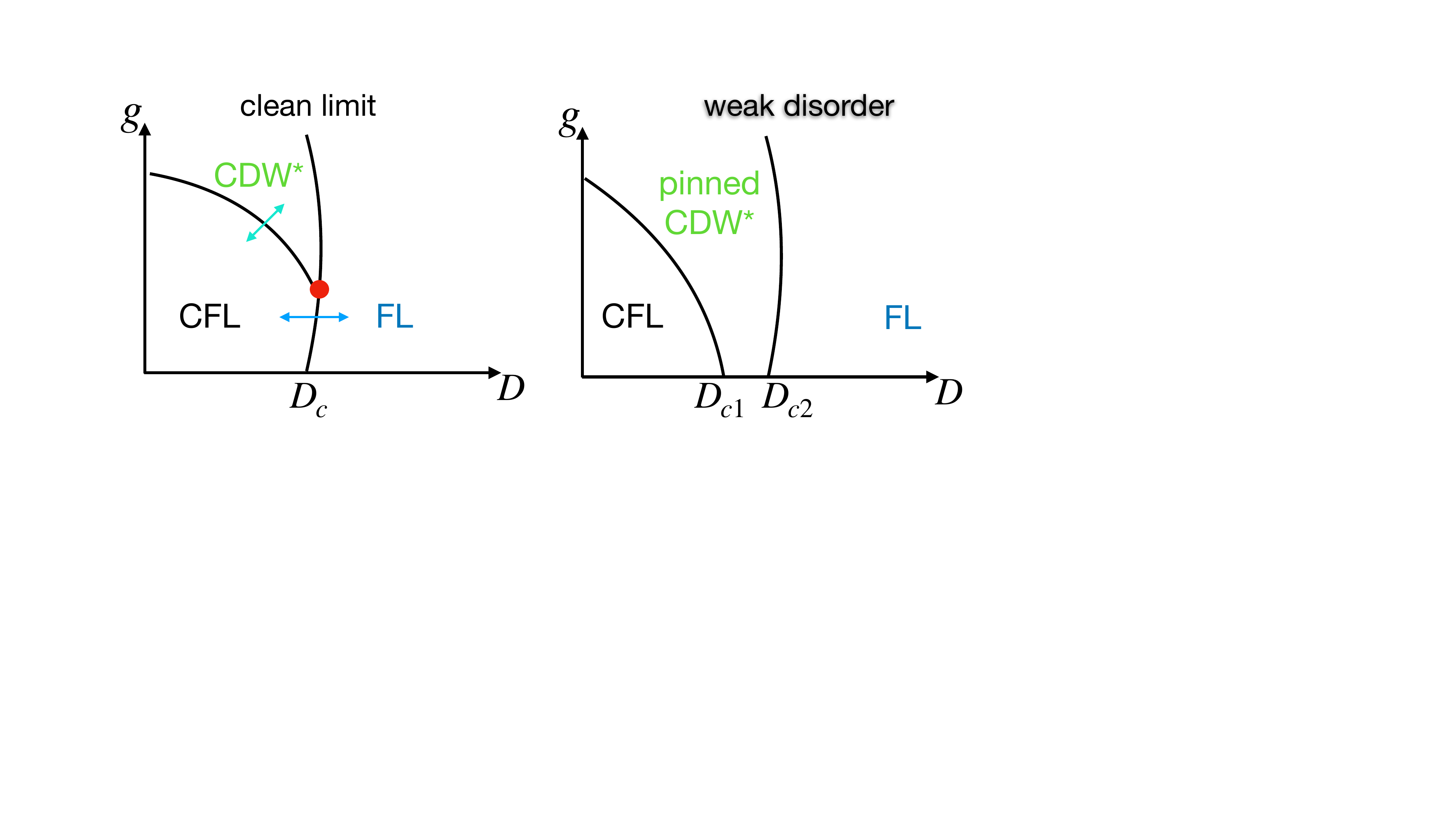}}
    \caption{A schematic phase diagram at $\nu = -\frac{1}{2}$. $D$ is the displacement field that tunes the CFL-FL transition. Left panel: the expected phases in a clean system. $g$ is some other parameter that allows access to the electrically insulating CDW$^*$ phase with long range charge order and a neutral fermi surface.  Right panel: Phase diagram with weak disorder: the (disordered version of ) CDW$^*$  will appear as an intermediate phase between CFL and FL. Long range charge order is lost in this phase but the diffusive neutral fermions will persist.  }
    \label{fig:phase}
\end{figure}

We show that both longitudinal and Hall resistivities have  universal jumps of order $\frac{h}{e^2}$ right at the quantum critical point. A similar effect was predicted for the continuous Mott transition in Ref. \onlinecite{senthil2008theory}. These universal resistivity jumps can provide a fingerprint of the putative continuous CFL-FL transition in a sufficiently clean sample. We consider the effects of disorder on the transition. We argue that weak disorder is very likely relevant and leads to a pinning of the fluctuating CDW order near the transition. The result is a novel intermediate phase between CFL and FL which is  electrically insulating but nevertheless has diffusive neutral fermionic excitations (ignoring localization effects on the composite fermions) coupled to a $U(1)$ gauge field. A schematic phase diagram sketching the pertinent phases is shown in Fig. \ref{fig:phase}.

In the clean limit, a theory of the evolution of the CFL to a CDW insulator has not been described in the literature. We present such a theory, at least in the case where the CDW has a coexisting neutral fermi surface coupled to a $U(1)$ gauge field (thus we will label the charge-ordered state CDW$^*$). This state is analogous to the Mott insulator with a coexisting spinon fermi surface discussed for Hubbard models on non-bipartite lattices. 

At $\nu = -3/4$, the CFL state is naturally described as the particle-hole conjugate of a composite fermi liquid at $1/4$ filling obtained by removing electrons from a filled Chern band. Thus we just focus on the nature of the CFL-FL transition at $1/4$ filling. A theory of this transition was left open in the previous literature. Following the same parton construction and logic as before requires first a theory of the Laughlin-superfluid transition at $1/4$ filling which is also not available in the prior literature. Here we will present a theory of this transition, and use it to construct a theory of the electronic CFL-FL transition at $1/4$ filling\footnote{ Our theory incorporates  lattice translation and filling constraints as crucial ingredients but does not respect $C_3$ symmetry. Thus it could potentially describe the CFL-FL evolution if the experimental system breaks $C_3$, as seems possible due to strain that is likely present. }. We also sketch an alternate possibility where the standard CFL at $1/4$ filling evolves to the FL through an intermediate CFL phase that is formed by binding 2 vortices to electrons. 

  The critical field theories discussed in this paper can describe phase transitions that are beyond the standard Landau paradigm for quantum criticality. The structure of these theories and their  construction are similar to those of deconfined quantum critical points\cite{senthil2004deconfined,senthil2004quantum,senthil2023deconfined} that have been discussed in various contexts before but mostly have not found experimental realization. Their possible realization in the moir\'e TMD context is thus an experimental opportunity to study a `beyond Landau' phase transition.

For other interesting work on phase transitions out of fractional Chern insulators in moir\'e systems, see Ref. \onlinecite{lee2018emergent}. A sample of early theoretical work   (which however did not incorporate filling constraints) on fractional quantum Hall phase transitions is in Refs. \onlinecite{kivelson1992global,wen1993transitions,chen1993mott}. 

 Very recently, FQAH states and possible a composite fermi liquid have been observed in a pentalayer graphene moir\'e system\cite{lu2023fractional}.  Our results will apply to this (or to any other future realizations) as well.

  \begin{figure}
\adjustbox{trim={.22\width} {.08\height} {.05\width} {.18\height},clip}
   {\includegraphics[width=.95\textwidth]{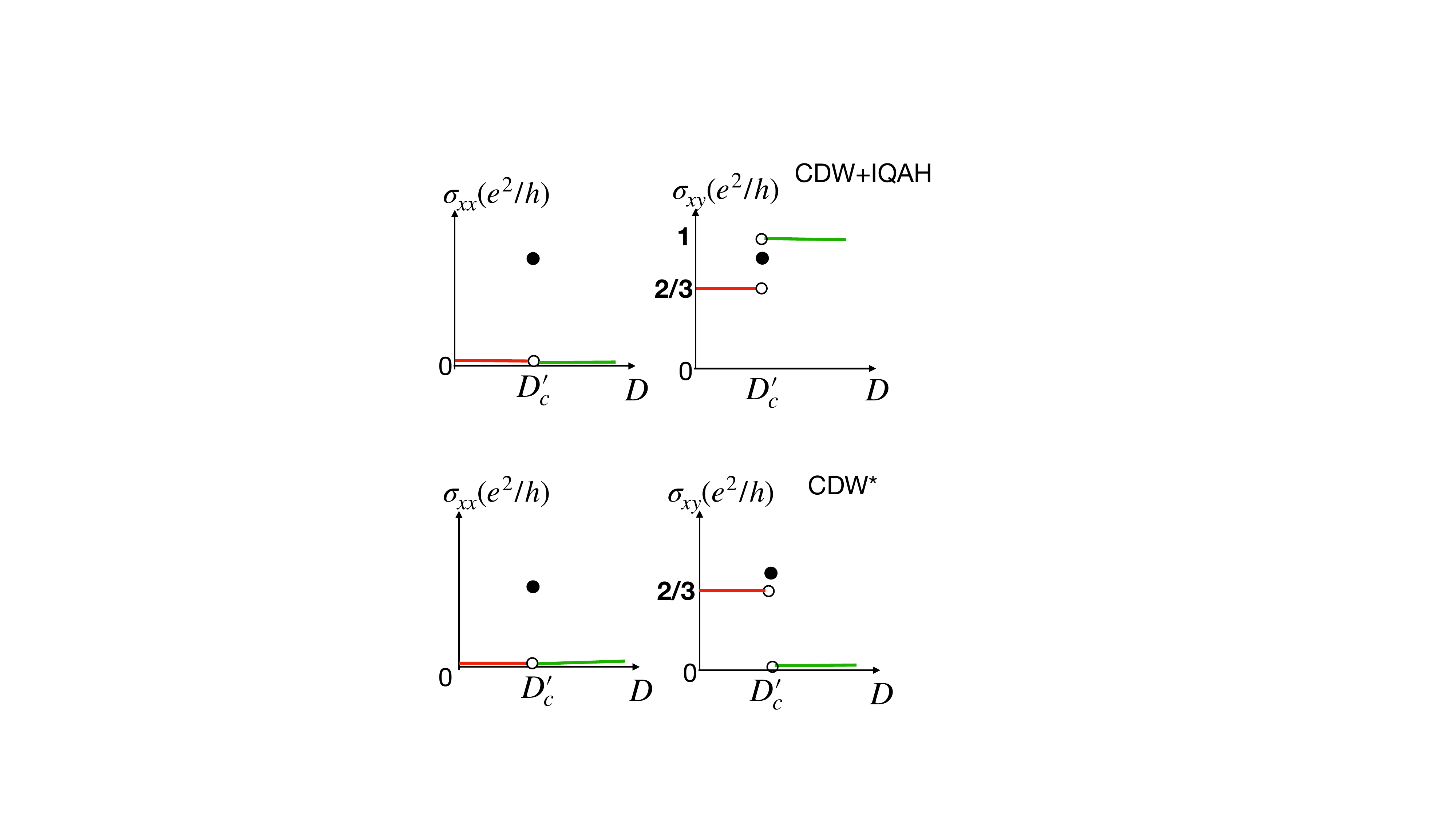}}
    \caption{ The conductivity across the FQAH-CDW transition at say, a filling of $2/3$.  The upper panel is  for the transition to CDW+IQAH (see Sec.  \ref{subsec:fqhtocdqiqah}). The lower panel is for the transition  to a CDW insulator with no electrical hall transport,  but with a neutral topological order (see Sec.  \ref{subsec:fqahtocdw*}). }
    \label{fig:rho_fqah}
\end{figure}

\begin{figure}
\adjustbox{trim={.18\width} {.44\height} {.1\width} {.24\height},clip}
   {\includegraphics[width=1.\textwidth]{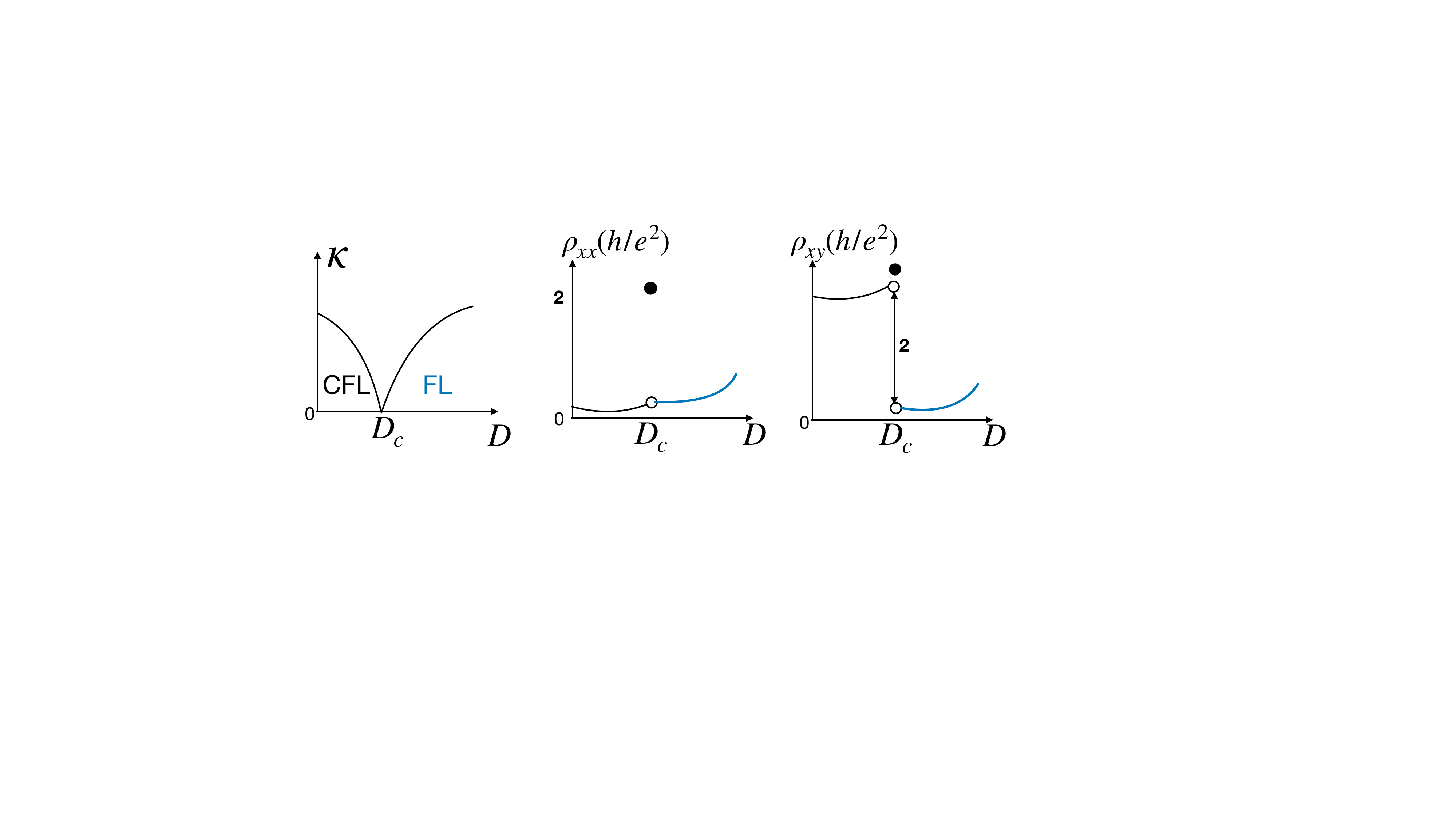}}
    \caption{The compressibility\cite{barkeshli2014continuous}, and resistivities across the CFL-FL (blue) transition at zero temperature in the clean limit. Both $\rho_{xx}$ and $\rho_{xy}$ have universal jumps at the critical point. The difference of $\rho_{xy}$ between the CFL and FL is $\frac{2h}{e^2}$, which is also approximately true even with  an intermediate `pinned CDW'$^*$ phase introduced by weak disorder.}
    \label{fig:observable}
\end{figure}

\begin{figure}
\adjustbox{trim={.22\width} {.3\height} {.05\width} {.24\height},clip}
   {\includegraphics[width=.85\textwidth]{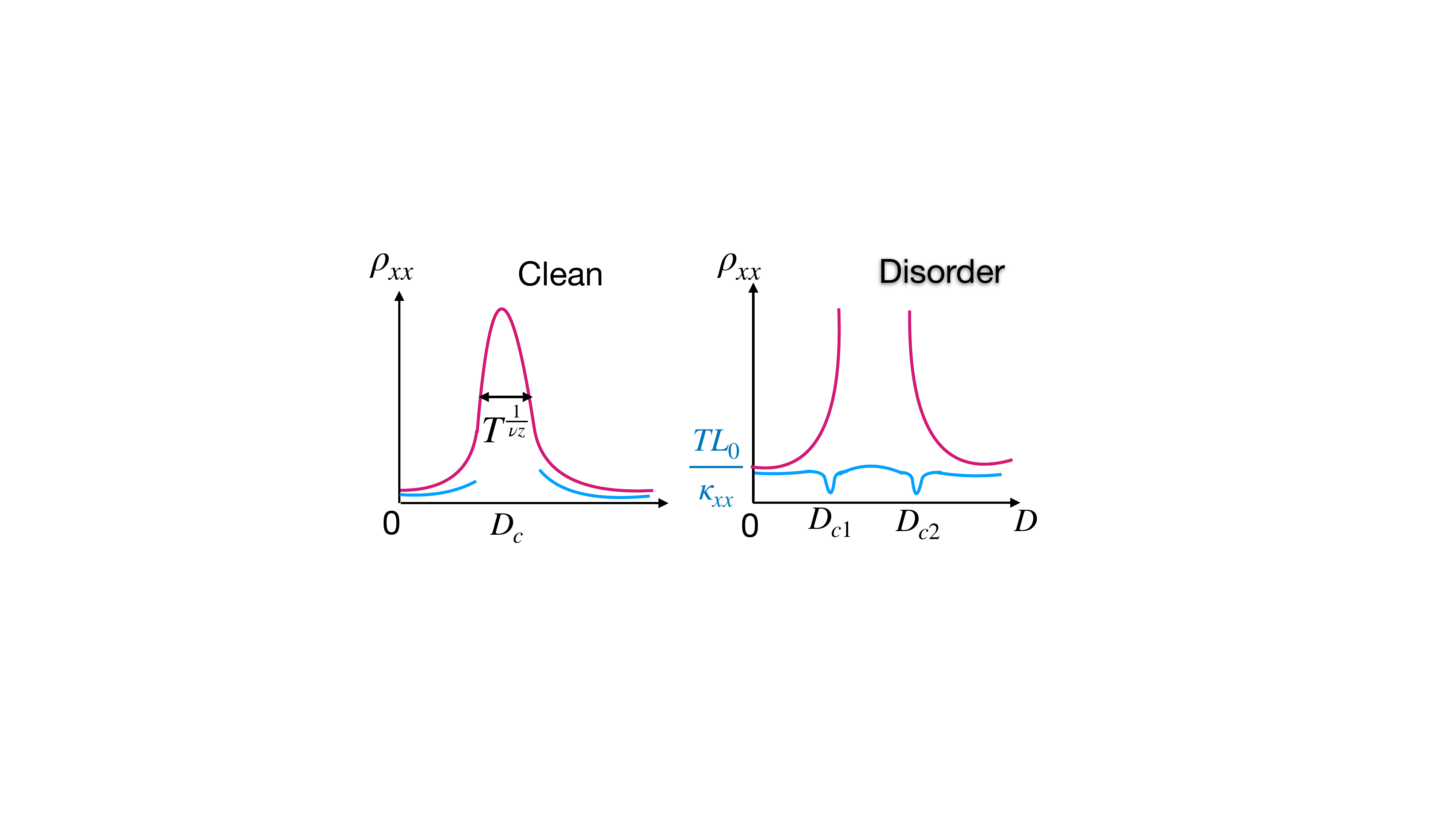}}
    \caption{The electrical and thermal resistivities across the CFL-FL transition, in the clean limit and with weak disorder, respectively. The blue line depicts the inverse thermal conductivity $TL_0/\kappa_{xx}$ which is proportional to the electrical resistivity inside both CFL and FL phases, but not near the critical point. $L_0=\pi^2k_B^2/(3e^2)$ is the Lorenz number. The intermediate CDW$^*$ phase is electrically insulating but has metallic thermal transport. There is a dip in $TL_0/\kappa_{xx}$ across the transition out of the intermediate CDW phase.}
    \label{fig:fT_rho}
\end{figure}

\begin{figure}
\adjustbox{trim={.08\width} {.3\height} {.01\width} {.2\height},clip}
   {\includegraphics[width=.7\textwidth]{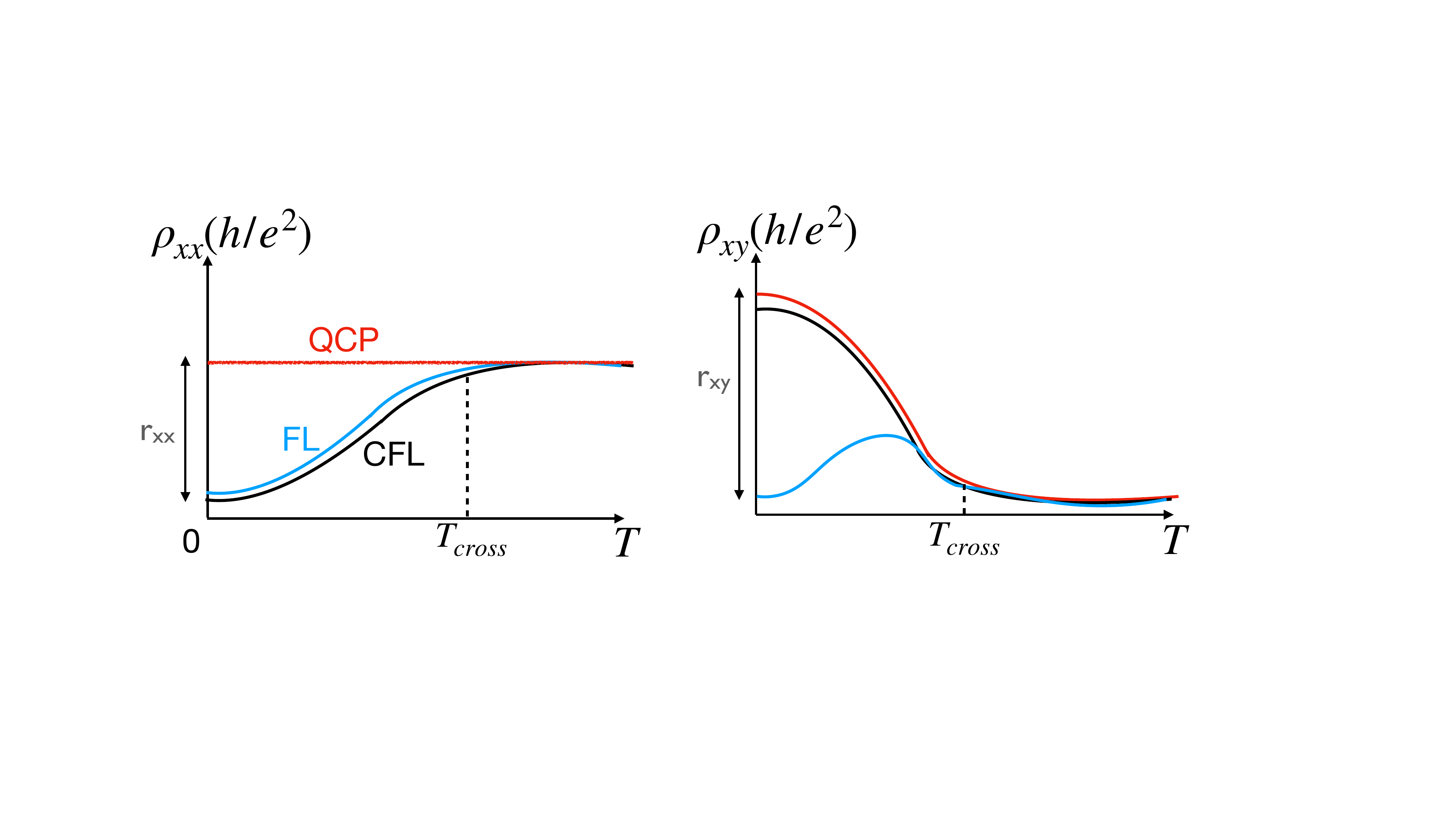}}
    \caption{The resistivity as a function of temperature in the CFL, FL and at the transition point, respectively. The blue curves are for the FL, the black for the CFL, and the red for the QCP. $r_{xx},r_{xy}$ are universal numbers of order $1$.}
    \label{fig:T_rho}
\end{figure}

\section{Preliminaries and summary of results} 
The tMoTe$_2$ (and other related moir\'e systems) have two valleys that the electrons can populate which are decoupled at the single particle level to an excellent approximation. In TMD systems like MoTe$_2$, the valley direction is locked strongly to the $z$-component of spin through spin-orbit coupling. 
Note that through out the region in parameter space where the FQAH (and other quantum Hall-like physics such as a possible CFL) is seen, and in the phases immediately proximate to it, ferromagnetism, as indicated by hysteretic transport near zero magnetic field, is observed. We will interpret this as valley polarization. In our theoretical treatment we will assume that the valley is maximally polarized and thus focus on the nature of the phase transitions that occur when only a single valley is occupied. Due to the spin-valley locking, electrons in a single valley can be regarded as spinless. Thus the theoretical problem we study is the nature of the quantum phase transitions out of quantum Hall phases of spinless electrons in a (partially filled) Chern band with Chern number $1$. 

Apart from the global $U(1)$ symmetry of charge conservation, the only other symmetries that may be present are lattice translation and $C_3$ rotations. Both these spatial symmetries are broken by quenched disorder while a uniform strain breaks $C_3$ but preserves translations.

\subsection{Summary of results}

Before going into details, we summarize our main results here. This will also provide an outline of the rest of the paper. 

\begin{enumerate}
    \item We first describe (Sec. \ref{sec:prtnfrmwrk}) a general framework for obtaining effective field theoretic descriptions for the phase transitions of interest. This framework is based on a parton construction and leads to field theories with emergent gauge fields coupled to gapless charged degrees of freedom, analogous to deconfined quantum critical points and other `beyond Landau' phase transitions. 
    \item
    We apply the parton framework to describe the intertwinement/competition between FQAH and CDW phases in terms of a continuum `quantum' Ginzburg-Landau theory. This theory is then used to study FQAH plateau transitions to CDW phases in Sec. \ref{sec:fqakpt}. Though our methods work at any rational lattice filling, we focus on filling $\nu=-\frac{2}{3}$ to  provide a critical theory between the FQAH phase and a CDW insulator phase.  A natural transition is to a CDW insulator with an integer quantum anomalous Hall effect. Crucially, if the CDW insulator has no integer quantum Hall effect, then within our formalism, it needs to be accompanied with a neutral topological order with a thermal Hall  effect. See Sec. \ref{subsec:fqhtocdqiqah} and \ref{subsec:fqahtocdw*} for details. 
    In Fig. \ref{fig:rho_fqah}, we sketch the expected behavior of the resistivities across either of these phase transitions. 
    
    \item At filling $\nu=-\frac{1}{2}$, we discuss the critical theory for the CFL to FL transition (see Sec. \ref{sec:1/2cflfl}).  Experimental signatures of the CFL-FL transition in transport and in compressibility are sketched in Figs. \ref{fig:observable},\ref{fig:fT_rho},\ref{fig:T_rho}.  In the clean limit, the compressibility vanishes at the critical point\cite{barkeshli2014continuous} while we show that the resistivites will have a universal jump.  We argue that there must be power law CDW fluctuations at the CFL-FL transition. We show  that weak disorder will pin the CDW orders and replace the CFL-FL critical point with an intermediate `pinned CDW*' phase (Fig. \ref{fig:phase}). Therefore an intermediate insulating phase always exist in the disordered sample.  Note that the `pinned CDW*' phase here is electrically an insulator, has no long range CDW order, but has mobile neutral diffusive fermions coupled to an emergent $U(1)$ gauge field.  
    
    \item We briefly describe an exotic CDW insulator (CDW$^*$) proximate to the CFL in a clean system, which has long-range CDW order coexisting with a neutral fermi surface (see Sec. \ref{sec:1/2cflcdw}). This phase can naturally be viewed as a parent that evolves into the `pinned CDW$^*$' with disorder. 
    
    \item At filling $\nu=-\frac{3}{4}$, we present a theory for the CFL to FL transition (see Sec. \ref{sec:3/4cflfl}) and the related problem of the  Laughlin-superfluid transition of bosons at $1/4$ filling. The latter is described by an interesting field theory with four massless Dirac fermions and three emergent $U(1)$ gauge fields. Lattice translation interchanges pairs of Dirac fermions and their associated gauge fields with each other. We also briefly discuss an alternate possibility where there is an evolution from the standard CFL (formed by electrons with four attached vortices) to CFL formed by electrons attached to two vortices to the standard FL (formed by electrons with no attached vortices). 
    
\end{enumerate}

\section{General parton framework for the phase transitions}
\label{sec:prtnfrmwrk}
The phase transitions of interest in this paper are conveniently accessed by a `parton' framework. Here we describe the ideas of the parton construction in general terms. We will repeatedly use this framework through out the paper. Suppose the electronic system of interest undergoes a phase transition out of some `base' phase P to other phases P'.  Suppose further that the effective field theory for phase P is  described by a Lagrangian that we write schematically as 
\begin{equation} 
{\cal L}_P = {\cal L}_P[c,\cdots; A]
\end{equation} 
Here $c$ is the electron operator and the ellipses represent other fields that may enter the low energy theory. For the specific case of incompressible quantum Hall phases, the electron itself is gapped, and the ellipses will stand for dynamical gauge fields with a Chern-Simons action. We have explicitly included an external  probe (background) $U(1)$ gauge field that is useful to keep track of the global $U(1)$ symmetry associated with electric charge conservation. 

We introduce a parton decomposition where the electron operator is written as 
\begin{equation}
    c(\mathbf r)=\Phi(\mathbf r) f(\mathbf r)
\end{equation}
This representation introduces a $U(1)$ gauge redundancy. Thus both $\Phi$ and $f$ will be coupled to a dynamical $U(1)$ gauge field\footnote{For experts, the background gauge field $A_\mu$ and the dynamical gauge field $a_\mu$ should both be viewed as spin$_c$ connections rather than ordinary $U(1)$ gauge fields. In contrast we will also encounter dynamical gauge fields which are ordinary $U(1)$ gauge fields. This is a useful bookkeeping device to keep track of whether objects charged under the gauge fields are bosons or fermions. For a review and references, see \cite{senthil2019duality}. We will not usually explicitly indicate this book-keeping scheme in our labelling of gauge fields in this paper though it will be implicit in our discussion.  } (which we denote $a_\mu$) with equal and opposite charges $\pm 1$.  The gauge redundancy is tied to a constraint that fixes the $\Phi$-density $\rho_\Phi$ to equal the $f$-density $\rho_f$. We can take the physical electric charge to either be carried by $\Phi$ or by $f$. For concreteness we assign it to $\Phi$.  Thus both $\Phi$ and $f$ will be at the same lattice filling as the electrons themselves: $\nu_\Phi = \nu_f = \nu_c$. Schematically we write the effective Lagrangian (on the lattice) in terms of the partons as 
\begin{equation} 
{\cal L} = {\cal L}[\Phi, A+a] + {\cal L}[f, -a]
\end{equation} 

Now we consider a `mean field' state ({\em i.e, } one where the fluctuations of the gauge field $a_\mu$ are turned off by hand)  where the $f$-fermions are themselves placed in the phase P of interest. The effective Lagrangian then takes the form 
\begin{equation} 
{\cal L} = {\cal L}[\Phi, A+a] + {\cal L}_P[f,\cdots; -a]
\end{equation} 
 Clearly if we condense the boson  $\Phi$  {\em i.e}, we place $\Phi$ in a `mean field' superfluid phase), then at long distances $a \approx -A$, and we recover the low energy Lagrangian for electrons in phase P. This is also obvious from the basic parton representation: once $\Phi$ condenses, we can identify $c$ and $f$, the dynamical gauge field is Higgsed, and the electrons are in phase P. 

A route to phase transitions out of P is to make the $\Phi$ field transition out of the `superfluid' phase. As an example, consider a transition (at some commensurate filling $\nu_\Phi = \nu_c$) into a boson Mott insulator (the precise nature of which will depend on the filling). This results in a new phase $P'$ of the electronic system. In the $P'$ phase, $\Phi$ can be integrated out and the low energy effective action of the full theory is given by 
\begin{equation}
    {\cal L}_{P'} = {\cal L}_P[f,\cdots; -a]
\end{equation}
 Note that the coupling to the probe gauge field $A$ has disappeared. Thus the phase $P'$ is an electrical insulator. The detailed low energy theory of P' is thus given by a gauged version of the low energy theory of the phase P. 

 If now we know an effective field theory description of the pertinent superfluid-Mott transitions of the boson $\Phi$ at its filling $\nu_\Phi$, then we have an effective field theory for the electronic phase transition out of phase P to phase P'. 

 An illustration of this parton framework is provided by the theory of the continuous Mott transition\cite{senthil2008theory} of spinful lattice electrons at half-filling between a Landau Fermi liquid (phase P) and 
a spin liquid Mott insulator (phase P') with a spinon Fermi surface. In this example  the low energy theory for the spinon Fermi surface is a gauged version of that of the electron Fermi liquid. The critical theory is the usual theory of the bosonic superfluid-Mott transition coupled (primarily through the gauge field $a$ but also by other terms) to the spinon Fermi surface. 

In this paper we are interested in situations where phase P is an electronic quantum Hall phase (either FQAH or CFL) at some fractional lattice filling. Then a natural bosonic Mott insulator for $\Phi$ is one that breaks lattice translation symmetry spontaneously, {\em i.e}, is a Charge Density Wave (CDW) insulator. Such bosonic superfluid-Mott transitions at fractional filling are conveniently described by passing to a dual vortex representation\cite{lannert2001quantum,balents2005putting,burkov2005superfluid}. At a rational filling $\nu = \frac{p}{q}$, the vortices now feel a mean magnetic flux of $\frac{2\pi p}{q}$ through each plaquette. Thus the action of lattice translations on the vortex fields is projective. With $2\pi p/q$ flux, there are q species of vortices which are related by translation action, and which hence will appear in the low energy effective theory. The corresponding Lagrangian for the boson sector takes the form 
\begin{equation}
 {\cal L}\left[\Phi_{vI}, b, a+A \right]  = \sum_{I = 1}^p |\left(\partial_\mu - i b_\mu \right)\Phi_{vI}|^2 +\cdots + \frac{1}{2\pi} b d(a+A) 
\end{equation}
Here $b_\mu$ is the usual dynamical $U(1)$ gauge field introduced by the particle-vortex duality. The ellipses represent terms in the effective action that will be constrained by the action of translation and any other space group symmetries present in the problem.  
The full effective theory that describes the phase transition out of phase P to the Mott insulator P' is then given by 
\begin{equation} 
{\cal L} = {\cal L}\left[\Phi_{vI}, b, a+A \right] +  {\cal L}_P[f,\cdots; -a]
\end{equation} 
If the $\Phi_{vI}$ are gapped (corresponding to the superfluid phase of $\Phi$), then we can do the integral over $b$ to set $a = -A$, and recover the phase P of the underlying electrons. Consider now the phase where the $\Phi_{vI}$ is condensed. The low energy long wavelength response to the probe $A$ field is then just given by a Maxwell action (with no Chern-Simons term). Thus we get an insulating state. However, the non-trivial translation action on $\Phi_{vI}$ ensures that translation symmetry is spontaneously broken. Thus we have a CDW insulator.  A theory of the  phase transition is then obtained by tuning to the critical point of the $\Phi_{vI}$ sector. 

Below we will use this general framework to discuss a variety of phase transitions out of quantum Hall phases. For an application of this framework to potentially continuous Wigner-Mott transitions of electrons at fractional filling from a Fermi liquid, see Ref. \onlinecite{musser2022theory} (see also Ref. \onlinecite{xu2022interaction} for work on related transitions). 

 Readers familiar with the theory of  deconfined quantum critical points will recognize the similarity with the structure of the theories we construct through this parton approach. This is perhaps not surprising since the transitions we seek to describe are inherently beyond the Landau symmetry breaking paradigm.

\section{Phase transitions out of the FQAH state}
\label{sec:fqakpt}
\subsection{Warm-up: FQAH to CDW transition at $\nu = \frac{1}{3}$}
\label{subsec:warmup} 
We begin by discussing a simple example of a phase transition associated with the destruction of the FQAH by studying electrons at a lattice filling $\nu_e = \frac{1}{3}$ in a band with Chern number $C = 1$. An FQAH with $\sigma_{xy} = \pm \frac{1}{3}$ is a possible ground state, and is characterized by a topological order.   A different possible insulating state is a CDW state that spontaneously breaks lattice symmetry. This state has no topological order. In current experiments on moir\'e systems, an FQAH at $\nu_e = \frac{1}{3}$ has not been observed. Nevertheless we begin by studying the FQAH-CDW transition at this filling as a warm-up which will help to introduce some of the ideas that will be used in the rest of the paper. 

At $\nu_e = \frac{1}{3}$, the FQAH state (the state P in the notation of the previous section) has an effective low energy Lagrangian 
\begin{equation} 
{\cal L}_P = -\frac{3}{4\pi} \alpha d\alpha + \frac{1}{2\pi} Ad\alpha 
\end{equation} 
where $\alpha$ is an ordinary dynamical $U(1)$ gauge field. 
 
A phase transition out of the quantum Hall phase can be thought of as a superfluid-insulator transition of the bosons\footnote{  At Laughlin fillings such as $1/3$, these bosons can be thought of as the `composite boson' (obtained by flux attachment) familiar from quantum hall physics in a continuum Landau level. The main novelty of what we are doing at these fillings is to incorporate filling constraints from the presence of the lattice. In the FQAH state itself, these constraints affect the translation action on the vortices of the composite boson. In subsequent sections, we will generalize the notion of composite bosons to handle Jain and other fractions.  } $\Phi$ at a lattice filling of $1/3$, and is described by a dual vortex theory in terms of fields $\Phi_{vI}$ with $I = 1, 2,3$. 
\begin{eqnarray}
\label{eq:1/3fqahcdw_lag}
    {\cal L} & = & {\cal L}\left[\Phi_{vI}, b \right]  -\frac{3}{4\pi} \alpha d\alpha -  \frac{1}{2\pi} ad\alpha    \\
    {\cal L}\left[\Phi_{vI}, b \right] & = &\sum_{I = 1}^3 |\left(\partial_\mu - i b _\mu \right)\Phi_{vI}|^2 +\cdots + \frac{1}{2\pi} b d(a+A) \nonumber
\end{eqnarray}
The ellipses represent terms consistent with the translation and $C_3$ symmetries of the lattice. 
The symmetry action of translations and $C_3$ (the subscript $I+1$ are defined modulo $3$,i.e. $I \mod 3 +1$)\cite{burkov2005superfluid} reads,
\begin{eqnarray}
    T_1&:& \Phi_{v,I}\rightarrow \Phi_{v,I+1}\nonumber\\
    T_2&:&\Phi_{v,I}\rightarrow e^{i\frac{2(I+1)\pi}{3}} \Phi_{v,I+1}\nonumber\\
    C_3&:&\Phi_{v,1}\rightarrow e^{i\frac{\pi}{6}} \Phi_{v,3},\Phi_{v,2}\rightarrow -i\Phi_{v,1},\nonumber\\
    &&\Phi_{v,3}\rightarrow e^{i\frac{\pi}{6}}\Phi_{v,2}.
    \end{eqnarray}
    The allowed quartic terms for the dual vortex fields read
\begin{eqnarray}
    \mathcal L_4[\Phi_{vI}]& = & {\cal L}_u + {\cal L}_v \nonumber \\
    {\cal L}_u & = & u(\sum_{I=1}^3 |\Phi_{vI}|^2)^2 \nonumber\\ 
    {\cal L}_v & = & v(|\Phi_{v1}|^2|\Phi_{v2}|^2+|\Phi_{v1}|^2|\Phi_{v3}|^2+|\Phi_{v2}|^2|\Phi_{v3}|^2), \nonumber
\end{eqnarray}
where the sign of $v$ determines the nature of the symmetry breaking state when $\Phi_{vI}$'s condense.

Now we can do the integral over the $a$-gauge field to set $b = \alpha$ to arrive at our final Lagrangian 
\begin{equation} 
\label{1/3fqhtocdw}
 {\cal L}  =  \sum_{I = 1}^3 |\left(\partial_\mu - i \alpha _\mu \right)\Phi_{vI}|^2 +{\mathcal L}_4 +\cdots  - \frac{3}{4\pi} \alpha d\alpha + \frac{1}{2\pi} Ad\alpha
\end{equation}

Clearly if the $\Phi_{vI}$ are all gapped, we get the Laughlin $1/3$ state. If however the $\Phi_v$ condense, then the $\alpha$-field is Higgsed and the quantum hall effect is destroyed. The low energy long wavelength response to the probe $A$ field is then just given by a Maxwell action (with no Chern-Simons term). Thus we get a CDW insulating state (where the non-trivial translation action on $\Phi_{vI}$ ensures that translation symmetry is spontaneously broken, and the absence of a Chern-Simons term for $A$ ensures absence of a Hall conductivity). The pattern of charge ordering in the CDW is determined by the structure of the non-linear terms  in Eqn. \ref{1/3fqhtocdw} (including a sixth order term that we have not explicitly shown)\cite{burkov2005superfluid}.

Thus the theory in Eqn.~\eqref{1/3fqhtocdw} enables discussion of the competition between the FQAH and CDW phases within a continuum field theory. It can thus be thought of as a quantum Ginzburg-Landau theory that is suitable for discussing universal aspects of the physics. In particular we can use it to study the phase transition between the topologically ordered FQAH state and a broken symmetry CDW insulator with no topological order.   

It should be emphasized that the field theory merely makes plausible a second order transition. Indeed a mean field treatment of the theory leads to a second order transition.  However the theory is strongly coupled and the mean field treatment will not be an accurate description of the true long wavelength physics. For instance we might expect that for $v > 0$ (when only one of the 3 $\Phi_{vI}$ bosons condense)  fluctuations may drive the transition first order while for $v < 0$ (when all 3 $\Phi_{vI}$ simultaneously condense) that it may be second order.  Whether it is actually second order or not beyond mean field  can only be determined through large scale numerical calculations that are beyond the scope of this paper. We therefore focus below on qualitative features of the transition that are implied by the structure of the field theory and that may be amenable to experimental study. 

The structure of the field theory implies scale invariant critical fluctuations with a dynamical exponent $z = 1$. Indeed we expect that the critical point will be described by a conformal field theory. As the putative critical point is approached, the charge gap $\Delta$ will vanish as a power law: 
\begin{equation} 
\Delta \sim |\delta|^{\bar{\nu} z} 
\end{equation} 
where $\delta$ is the parameter tuning the transition at zero temperature (with $\delta = 0$ chosen to be the location of the criitcal point). In the moir\'e TMD context, this is usually the deviation of the displacement field from its critical value, {\em i.e} $\delta = D - D_c$. The exponent $\bar{\nu}$ is not known though it has the value $1/2$ within mean field theory.

Why does the charge gap vanish though the transition is between two insulators? The reason is that the FQAH state has a non-zero Hall conductivity which vanishes in the CDW insulator. Thus the charge physics is intimately involved in the criticality. The electronic compressibility $\kappa = \frac{dn}{d\mu}$ vanishes exponentially in either phase at temperatures $T$ below the charge gap. However at the critical point (from standard scaling arguments in space dimension $d = 2$ with $z = 1$) it vanishes as a power-law 
\begin{equation}
    \kappa \sim T 
\end{equation}

Furthermore the critical theory will have universal electrical conductivities $\sigma_{xx}$ and $\sigma_{xy}$ at $T = 0$. Thus (much like in other quantum Hall plateau transitions discussed in the literature\cite{sondhi1997continuous}) the longitudinal conductivity will jump to a non-zero value just at the critical point. At a non-zero but low temperature, this jump will be rounded in a universal manner. Indeed as a function of $\delta$ and $T$, the conductivities will satisfy a universal scaling relation 
\begin{equation} 
\label{eq:condscal}
\sigma_{ij}(\delta, T) = \frac{e^2}{h} S_{ij}\left( \frac{\delta}{T^{\frac{1}{\bar{\nu} z}}} \right)  
\end{equation}
where $S_{ij}(x)$ is a universal $2 \times 2$ matrix. The longitudinal component $S_{xx} = S_{yy}$ satisfies 
\begin{eqnarray}
    S_{xx}(x \rightarrow 0) & \rightarrow & c_{L} \\
    S_{xx} (x \rightarrow \pm \infty) & \rightarrow & 0 
    \end{eqnarray} 
    where $c_L$ is a non-zero constant of order $1$. 
    The Hall component $S_{xy}$ will similarly satisfy 
    \begin{eqnarray} 
 S_{xy}(x \rightarrow 0) & \rightarrow & c_{H} \\
    S_{xy} (x \rightarrow \infty) & \rightarrow & 0 \\
     S_{xy} (x \rightarrow - \infty) & \rightarrow  & \frac{1}{3} \\
    \end{eqnarray} 
 $c_H$ is also a constant of order $1$. (We have taken $\delta < 0$ to be the FQAH phase). In both the longitudinal and Hall channels, the conductivities will approach their zero temperature values in either phase exponentially as $T$ is decreased. 

 The transport features and the compressibility are good targets as experimental probes of the transition in the near term. More challenging will be to extract the critical behavior of the order parameters for the charge ordering. These 
 will have power law correlations with a universal exponent: 
\begin{equation} 
\langle O_{CDW}(\vec x) O_{CDW}(\vec x') \rangle \sim \frac{1}{|\vec x - \vec x'|^{2\Delta_{CDW}}}
\end{equation} 
The `scaling dimension' $\Delta_{CDW}$ is not known. The mean field estimate is $\Delta_{CDW} = 1$. We expect that gauge fluctuations beyond mean field will reduce the scaling dimension: this is because the CDW order parameter $O_{CDW}$ is a particle-hole bilinear of the basic fields $\Phi_{vI}$; the gauge field leads to an attractive interaction between particles and holes and thus $O_{CDW}$ will have slower correlations than in  mean field theory. We therefore expect that the true $\Delta_{CDW} < 1$. The electronic tunneling density of states will also have a power-law singularity at the critical point. The $T = 0$ tunneling conductance $G(V)$ will satisfy 
\begin{equation}
    G(V) \sim |V|^y
\end{equation}
where $y$ is a universal critical exponent. Note that $G(V)$ will be zero at low bias voltages in either phase. The exponent $y$ is related to the electron scaling dimension $\Delta_e$ through $y = 2\Delta_e - 1$. Though the precise value is not known, we expect that $\Delta_e$ is at least larger\footnote{The electron operator is a product of a monopole operator of $\alpha$ and three $\phi_v$ operators. Though the monopole scaling dimension is not known, each $\phi_v$ has a mean field scaling dimension of $1/2$.  Hence we might expect $\Delta_e$  is at least equal to $\frac{3}{2}$)} than $1$  so that the tunneling conductance will be sublinear. Nevertheless the presence of power law conductance at the transition between two gapped phases underscores the importance of the charge degrees of freedom at the critical point.

The discussion so far is for an ideal clean system. How does the physics change upon including the presence of weak quenched disorder? There is a linear coupling of the random potential to $O_{CDW}$: 
\begin{equation}
    \int d^2x d\tau V(x) O_{CDW}(x) 
    \label{eq:dis_cdw}
\end{equation}
where $V(x)$ may be taken to be a random Gaussian variable with a variance $g$. The FQAH  phase is of course stable to disorder; however the linear coupling to disorder implies that  long range CDW order will be destroyed at the longest scales owing to the instability of ordering in random field systems in two dimensions. A sharp transition associated with the destruction of the topological order will nevertheless still be present. Furthermore even if the clean transition is first order, it will be rendered continuous at the longest scales in the presence of disorder. 

Disorder is likely relevant at the putative clean quantum critical point.  Standard scaling arguments show that the coupling in eq \eqref{eq:dis_cdw} is relevant if the scaling dimension $\Delta_{CDW} < 2$. As we have already argued that very likely $\Delta_{CDW} < 1$, it follows that disorder is relevant. 
The effect of long range Coulomb interactions can be included into the theory at least at a qualitative level, as discussed in the literature\cite{fisher1990coulomb}.  In the presence of both disorder and the long range part of the Coulomb interaction, the transition out of the FQAH state into the pinned CDW will have dynamical exponent $z = 1$. The universal transport near the critical point described above will continue to hold (although the detailed form of the scaling function will be modified).

\subsection{FQAH to CDW-IQAH state at $\nu = -\frac{2}{3}$}
\label{subsec:fqhtocdqiqah}
The FQAH  seen at $\nu = -\frac{2}{3}$ in tMoTe$_2$ can be understood by starting with the $\nu = -1$ state, and  doping electrons to reach $\nu = -\frac{2}{3}$. The electrons are at a filling $\nu_e = \frac{1}{3}$ and form an FQAH state. Thus, we can view the observed FQAH state at $\nu = - \frac{2}{3}$ as the Integer Anomalous Quantum Hall (IAQH)  state of holes at $\nu = -1$, coexisting with an FQAH state of electrons at $\nu_e = \frac{1}{3}$. 

Using the results of the previous subsection, we can now easily describe a phase transition where the $\nu = -\frac{2}{3}$ FQAH evolves into a CDW insulator coexisting with an IQAH with $\sigma_{xy} = \pm 1$. (Simply consider the $\nu_e = \frac{1}{3}$-CDW transition in the background of the $\nu = -1$ IQAH phase). Clearly the universal critical properties are the same as described in the previous section, except that the total physical conductivity tensor  will be given by 
\begin{equation}
    \sigma_{ij} = \sigma_{ij}|_{\nu = 1/3} + \sigma_{ij}|_{\nu = -1} 
    \end{equation} 
    where $\sigma_{xy}|_{\nu = 1/3}$ evolves across the transition as described in the previous subsection. The IAQH on the other hand has the familiar $T = 0$ conductivities $\sigma_{xx}|_{\nu = -1} = 0, \sigma_{xy}|_{\nu = -1} = \pm \frac{e^2}{h}$.


\subsection{FQAH -CDW* transition at $\nu = -\frac{2}{3}$}
\label{subsec:fqahtocdw*}
We now describe a phase transition between the FQAH state at $\nu = -\frac{2}{3}$ and a CDW state with no quantum Hall electrical transport. We begin by recalling the Chern-Simons effective theory of the $\frac{2}{3}$ FQAH state (which is our state P in this subsection) with the action 
\begin{equation}
\label{2/3cstheory}
    {\cal L}_P = -\frac{1}{4\pi} \alpha d\alpha + \frac{3}{4\pi} \beta d\beta + \frac{1}{2\pi} A d(\alpha + \beta) 
\end{equation}
Here $\alpha, \beta$ are (ordinary) dynamical $U(1)$ gauge fields while $A$ is an external probe gauge field. The $\alpha$ field describes the filled Chern band while the $\beta$ field describes the $1/3$ FQAH of the doped electrons. 

We follow the parton strategy discusse above and put $f$ in the $2/3$ FQAH state described by the Chern-Simons action of Eqn.~\eqref{2/3cstheory}  and consider the superfluid-CDW insulator transition of the $\Phi$ field at $2/3$ filling. 
  The vortices in $\Phi$ now see a flux $4\pi/3$ through each lattice plaquette, and the resulting projective action of translations results in 3 species of vortex fields $\Phi_{vI}$ ($I= 1,2,3$) that appear in the low energy theory which therefore takes the form
\begin{align} 
{\cal L} &= \sum_{I = 1}^3{\cal L}[\Phi_{vI} , b] + \frac{1}{2\pi} bd(A + a)  -\frac{1}{4\pi} \alpha d\alpha \notag \\
&~~~+ \frac{3}{4\pi} \beta d\beta - \frac{1}{2\pi} a d(\alpha + \beta) 
\end{align} 
 If the vortices $\Phi_{vI}$ are all gapped (corresponding to condensation of $\Phi$), then we can set $a = -A$ at long distances to get the electronic $2/3$ FQAH state. Consider a different phase where $\Phi_{vI}$ is condensed. As in our warm-up example, this leads to CDW order. Further, the $\Phi_{vI}$ condensation sets $b = 0$. The coupling of the $A$ field to the remaining fields thus vanishes, and we get an electrical insulator with no quantized Hall conductivity. 

The remaining fields $\alpha, \beta, a$ have the action 
\begin{equation}
  -\frac{1}{4\pi} \alpha d\alpha + \frac{3}{4\pi} \beta d\beta - \frac{1}{2\pi} a d(\alpha + \beta)    
\end{equation}
We can now integrate $a$ out to set $\alpha = -\beta$. The result is a theory with the action 
\begin{equation} 
\frac{2}{4\pi} \beta d\beta
\end{equation} 
This describes a topological order (denoted $U(1)_2$) in the {\em electrically neutral} sector of the theory. There is a single anyon (a neutral semion).   

Thus the CDW state we have reached has coexisting topological order characterized by a neutral semion. We thus categorize this state as CDW$^*$ to distinguish from the conventional CDW. However this CDW$^*$ will have no distinction in electrical transport from the conventional CDW. It will however have a quantized  thermal Hall effect (and associated neutral edge currents) which distinguish it from the standard CDW.  

The physical properties of the transition and their scaling structure will be similar to what we described above in our warm-up example (though the specific critical exponents will be different). The inequality $\Delta_{CDW} < 1$  is expected to hold at this transition as well; thus the remarks on relevance of disorder will also apply. 

Could there be a direct transition from the $2/3$ FQAH state to a conventional CDW insulator with no quantum Hall effect? While we do not currently have a theory of such a transition, we emphasize that it might be interesting to note that the insulating state found near the FQAH state in experiments may actually have a `dark' topological order not visible to the standard electrical transport probes. Fig \ref{fig:rho_fqah} schematically plots the conductivity behavior across the two FQAH-CDW transitions discussed above.

\section{CFL to FL transition}
\label{sec:1/2cflfl}
Now we  move on to the  theory of a continuous transition from CFL to FL. This transition was first discussed in  Ref.  \onlinecite{barkeshli2014continuous}. Below we will  build on and further  refine this theory.   Crucially we will incorporate filling constraints coming from fractional filling of the lattice into the low energy theory, as is appropriate for any low energy effective theory. We will show that these constraints modify the physics, and lead naturally to the possibility of a direct second order CFL-FL transition protected just by translation symmetry, unlike the situation considered in Ref. \onlinecite{barkeshli2014continuous}.   We will use the resulting understanding to extract testable predictions for experiments on moir\'e TMD systems. The parton representation $c = \Phi f$ continues to provide a unified framework to describe both the CFL and the FL and their phase transition.   In this section we focus on the filling $\nu_c=\frac{1}{2}$  per unit cell on average. As explained earlier this implies that $\Phi$ and $f$ are also at lattice filling $1/2$.  We choose the Fermi liquid as our `base' phase P.  As before this can be viewed as placing $f$ in a `mean field' Fermi liquid state and condensing the boson $\Phi$. 
To describe the CFL state, we keep the $f$-fermions in their `mean field' fermi liquid state, but take the $\Phi$ bosons to form a $\nu = 1/2$ Laughlin state.  In the next section we will consider an alternate situation where the bosons form a CDW Mott insulator, which will correspond to an exotic CDW phase of the physical electrons.  In this section we focus on the CFL to FL transition.  

More concretely, the effective low energy theory of the physical electron can be  written:

\begin{align}
S_c=S_b+S_f +S_{bf} 
\label{eq:CFL_FL_whole}
\end{align}
where
\begin{align}
S_f&=\int dt d^2x f^\dagger(t,\mathbf x)(i\partial_t +i a_0+\mu)f(t,\mathbf x) \notag \\ 
&~~~+\frac{\hbar^2}{2m}f^\dagger(t,\mathbf x) (\mathcal{\nabla}+i\mathbf a)^2 f(t,\mathbf x)
\label{eq:CFL_FL_Sf}
\end{align}
describes a state with a Fermi surface at filling $n_f=\frac{1}{2}$ per unit cell. The $S_b$ describes the dynamics of $\Phi$ as it  undergoes a bosonic $\nu = 1/2$ Laughlin-superfluid transition. As usual  $\Phi$ couples to $a + A $ (we remind that $A$ is a probe gauge field, and $a$ is a dynamical gauge field), and $f$ to $a$. The $S_{bf}$ represents couplings between the $\Phi$ and $f$ sectors which we  will discuss below.

Questions closely related to the $\nu = 1/2$ FQHE to SF transition of bosons have been discussed in previous papers, and we briefly adapt (and refine)   the results as needed here. There are in fact multiple equivalent Lagrangian descriptions\cite{song2022deconfined} of this transition which are related to each other through duality transformations.   For concreteness, we work with  just one of the equivalent theories which is based on a further parton decomposition of $\Phi$ as a product of two fermions: 
\begin{equation}
\Phi = f_1 f_2
\end{equation}
In this parton representation, the fermions $f_1$ and $f_2$ are both at filling $1/2$. There is a $U(1)$ gauge field\footnote{Actually a spin$_c$ connection. Also strictly speaking this parton representation introduces an $SU(2)$ gauge redundancy but we assume that the mean field states of interest break the $SU(2)$ gauge structure to $U(1)$} that we will denote $\hat{a}_\mu$. Let us temporarily treat the gauge field $a + A$ that couples to $\Phi$ as a background field and denote it by $A_b$. We assign the charge of $A_b$ to $f_1$ so that $f_2$ is neutral under $A_b$.  We take a mean field state where $f_1$ and $f_2$ each see $\pi$ flux through each plaquette. This doubles the unit cell for the band structure of both $f_1$ and $f_2$. Thus at half-filling, they may both form band insulators. We will assume that these `mean field' band insulators  have Chern numbers $C_1$ and $C_2$ for $f_1$ and $f_2$ respectively. 

Consider a mean field state  where $C_1 =  1, C_2 = - 1$.   As argued in Ref. \onlinecite{barkeshli2012continuous} this is a superfluid phase of $\Phi$. On the other hand, if $C_1 = 1, C_2 = 1$, we get the $\nu = 1/2$ FQH state of bosons. Thus the bosonic $1/2$ FQHE to superfluid transition occurs through a change of Chern number of $f_2$ from $C_2 = -1$ to $C_2 = +1$ while $C_1$ stays fixed at $1$. The Chern number changing phase transition of $f_2$ can occur through a band touching which can be described (generically) in terms of two massless Dirac fermions. These Dirac fermions will be coupled to the $U(1)$ gauge field $\hat{a}$, and will have a background Chern-Simons term coming from the filled $f_1$ band. The resulting field theory thus takes the form 
\begin{align}
\mathcal L&=\bar \psi(\gamma_\mu (-i \partial_\mu \sigma_0 -\hat{a}_\mu \sigma_0 )) \psi+m\bar\psi\psi \notag \\ 
&~~~+\frac{1}{4\pi}d(\hat{a} + A_b) (\hat{a} + A_b) - \frac{1}{4\pi} \hat{a} d \hat{a}   
\label{eq:U1_2psi}
\end{align}
where we have two Dirac fermions $\psi_1, \psi_2$ with $\sigma_a$ as the Pauli matrix in the flavor space.  The transition is tuned by the Dirac mass term $m\bar{\psi}{\psi}$. (We have defined a single massless Dirac fermion through a Pauli-Villars scheme with another heavy Dirac fermion in the UV. Then for $m < 0$ a single Dirac fermion has a Hall conductivity $0$ and for $m > 0$ a Hall conductivity $+1$.) 

What guarantees that the $f_2$ fermions change their Chern number by $2$ across the transition and not by $1$ as usually happens in band theory? The answer is that due to the $\pi$ flux seen by $f_2$ on the lattice, translations act projectively on them. This leads to a doubling of fermion species in the low energy theory. Thus if there are Dirac nodes, we will have an even number of them, and the generic situation is when we have two Dirac nodes. 
Note that as written there is a `flavor' $SO(3)$ symmetry that involves rotations in the $\psi_1, \psi_2$ space. 
One worry of Ref.~\onlinecite{barkeshli2014continuous} is that there are three corresponding `mass'  terms  $\bar \psi \sigma_{a} \psi$ with $a=1,2,3$, which may split the direct transition. Ref.~\onlinecite{barkeshli2014continuous} mentioned that these terms may be forbidden by crystal symmetry, but no concrete argument was provided.  Here we see that such terms must be forbidden by the action of translation invariance on the $f_2$. Ultimately this action arises from combination of translation from the filling constraint of the bosons on the translation invariant lattice\footnote{To understand this directly, note that  if we add $\bar \psi \sigma_a \psi$ at the critical point $m=0$, we reach a trivial insulator. However, no symmetric trivial insulator is possible at half filling unless translation symmetry is broken. Thus $\bar \psi \sigma_a \psi$ must break the translation symmetry and correspond to CDW order parameters.  Thus the symmetry properties of the operators in the critical theory are fixed by the LSM constraint so that no microscopic calculations are needed.}  It follows that these flavor mass terms $\bar \psi \sigma_{a} \psi$ with $a=1,2,3$ must transform under translations, and hence must correspond to CDW order parameters.  In the Dirac theory, it is natural to expect that the quartic terms are irrelevant\footnote{This is definitely true in the free Dirac theory without any gauge field} so that there is an emergent SO(3) symmetry rotating the three CDW orders $O^a_{CDW} =\bar \psi \sigma_a \psi$. This conclusion does not rely on the microscopic $C_3$ rotation symmetry, implying that the $C_3$ breaking term is irrelevant at this fixed point with emergent SO(3) symmetry. In Appendix \ref{app:dual}  we argue that the 3 CDW order parameters are associated with momenta $\mathbf M_1, \mathbf M_2, \mathbf M_3$ in the hexagonal Brillouin zone. We also note that there are also two other equivalent dual theories of this phase transition (see Appendix.\ref{app:dual}), as first pointed out by two of us in the context of chiral spin liquid to superconductor and CDW transition\cite{song2022deconfined}.

In passing we note that it has been suggested\cite{barkeshli2015continuous}  that the second order Laughlin-superfluid transition of bosons at half-filling requires inversion symmetry based on a different field theoretic construction.   However, as we just demonstrated, inversion (which is not present in the moir\'e system of interest) is not necessary. Translation and filling constraints alone are enough to protect a direct transition. 

  \begin{figure}
  \adjustbox{trim={.08\width} {.48\height} {.14\width} {.2\height},clip}
   { \includegraphics[width=1.\textwidth]{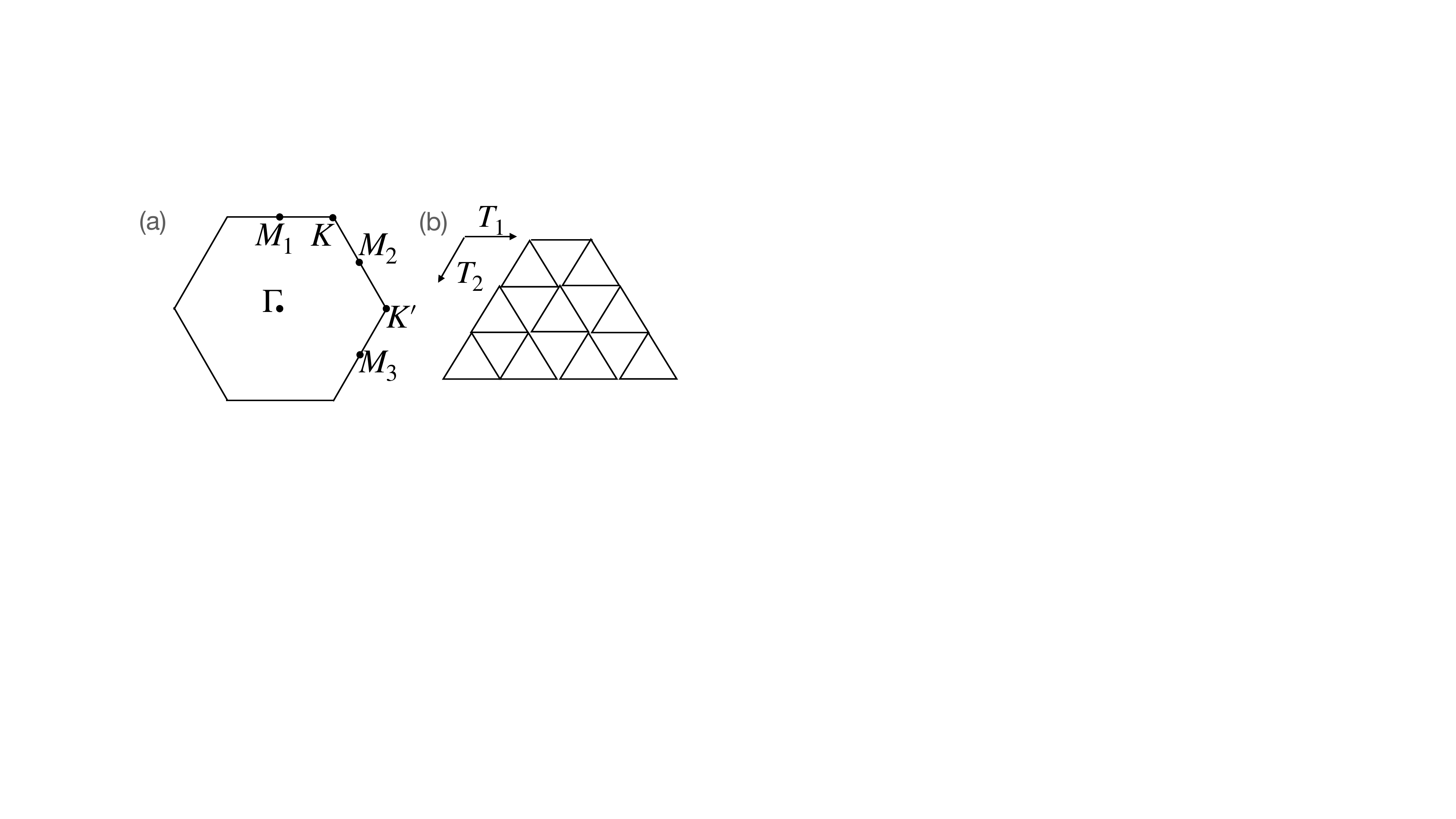}}
    \caption{The triangular lattice and the Brillouin zone. It illustrates $T_1, T_2$, momentum $\Gamma, M_1, M_2, M_3, K, K'$. }
    \label{fig:triangle}
\end{figure}

With this understanding of the building block of the superfluid-FQHE transition of bosons at $\nu = 1/2$, let us return to the full electronic theory of Eqn.~\eqref{eq:CFL_FL_whole}. The coupling of this sector to the $f$-fermi surface was also discussed in Ref. \onlinecite{barkeshli2014continuous} following the arguments of Ref. \onlinecite{senthil2008theory}. At $T = 0$, the gauge fluctuations are Landau damped by the fermions and this can be shown to lead to them decoupling  from the critical boson sector. It remains to understand the effect of other couplings between local operators in the boson and fermion sector.


First there will be a coupling of `energy densities'  of the form  $S_{bf}=\int dt d^2x  |\Phi|^2 f^\dagger f$. Integrating $f$ gives a perturbation to the boson action $\delta S_b=A \int |\frac{\omega}{q}| |O(\omega,q)|^2$ with $O=\Phi^\dagger \Phi$. This perturbation is irrelevant when $\bar{\nu}>\frac{2}{3}$ where $\bar{\nu}$ is the correlation length exponent (not to be confused with the filling) for the FQHE-SF critical point in the boson sector.  Although there is no precise estimation of  the exponent for the FQHE-SF transition so far, a large N calculation using the Dirac theory suggests that $\bar{\nu} >\frac{2}{3}$ likely holds\cite{barkeshli2012continuous}. In any case we will assume this to be true, and treat this as a bound on the exponent that has to be satisfied for the theory to apply. 

Second, there will be a coupling of the CDW order parameters $O^a_{CDW} = \bar \psi \sigma_{a} \psi$ to the Fermi surface (which was not considered in Ref. \onlinecite{barkeshli2014continuous}).  This will lead to a Landau damping term 
\begin{equation}
\int d^2\vec q d\omega |\omega| |O^a_{CDW} (\vec q, \omega)|^2 
\end{equation} 
This will be irrelevant at the bosonic quantum critical point  if the $O^a_{CDW}$ scaling dimension  $\Delta_{CDW}  > 1$. In the free massless Dirac theory, these operators have scaling dimension $2$. It is expected that the scaling dimension will be reduced by gauge fluctuations so that at the true fixed point we will have $\Delta_{CDW} < 2$. It is not currently known what the true scaling dimension is but for the theory of Ref. \onlinecite{barkeshli2014continuous} to really apply, it is necessary that $\Delta_{CDW}  > 1$ and we will assume this is true. 

Thus, provided the  two bounds on exponents $\bar{\nu} > 2/3, \Delta_{CDW} > 1$ are satisfied,  the bosons $\Phi$ and fermions $f$ dynamically decouple at the CFL-FL transition. 
The physics of the critical boson sector is then just described by the $\nu = 1/2$ FQHE to SF transition described above, which is presumably described by a conformal field theory (CFT). The dynamical critical exponent is thus $z = 1$. We show in appendix \ref{app:density} that the flux of $a$ will not be spontaneously generated at the CFL-FL critical point, hence the continuous transition will be preserved.
 
In the following we point out some major experimental signatures for this transition.








\subsection{Resistivity and compressibility around the CFL to FL transition}
\label{subsection:resistivity_cfl_fl}
Here we discuss the transport response at the QCP in the clean limit. Using the Ioffe-larkin rule, the resistivity tensor of the physical electron is

\begin{equation}
    \rho_c=\rho_\Phi+\rho_f
    \label{eq:ioffe_larkin_rho}
\end{equation}
where $\rho_a=\begin{pmatrix} \rho_{a;xx} & \rho_{a;xy} \\ \rho_{a;yx} & \rho_{a;yy}\end{pmatrix}$ is a $2\times 2$ tensor.

We expect $\rho_f$ to be smooth across the transition and is given by 
\begin{equation}
    \rho_f=\begin{pmatrix} \rho_f & \rho_{f;xy} \\ -\rho_{f;xy} & \rho_f \end{pmatrix}
\end{equation}
where $\rho_f$ and $\rho_{f;xy}$ are non-universal numbers.  We expect $\rho_f \ll \frac{h}{e^2}$  when the disorder strength in the system is not very large. The Hall conductivity $\sigma^f_{xy}$ will be determined by the Berry curvature enclosed by the Fermi surface, and will be some fraction of $\frac{e^2}{h}$. If $\sigma^f_{xx}$ is large compared to $\frac{e^2}{h}$ (as expected in the weak disorder limit), then we expect that $\rho_{f;xy} \approx \frac{\sigma^f_{xy}}{\left(\sigma^f_{xx}\right)^2}$ is small.

On the other hand, the resistivity tensor for the boson $\rho_\Phi$ goes through a jump across the transition.  In the CFL phase, we have $\rho_\Phi=\begin{pmatrix}0 & -2 \\ 2 & 0 \end{pmatrix} \frac{h}{e^2}$.  In the FL phase, we have $\rho_\Phi=\begin{pmatrix} 0 & 0 \\ 0 & 0 \end{pmatrix}\frac{h}{e^2}$. Exactly at the critical point, the CFT nature of the $S_b$ leads to a universal tensor $\rho_\Phi=\frac{1}{\sigma_\Phi^2+\sigma_{\Phi;xy}^2} \begin{pmatrix} \sigma_\Phi & -\sigma_{\Phi;xy} \\ \sigma_{\Phi;xy} & \sigma_\Phi \end{pmatrix}$ where $\sigma_\Phi$ and $\sigma_{\Phi;xy}$ are universal numbers at order of $\frac{e^2}{h}$.  Actually as discussed for the closely related problem of the continuous Mott transition\cite{witczak2012universal}, if the QCP is approached by cooling from non-zero $T$, the gauge fluctuations do affect the bosonic criticality. This is because the zero Matsubara frequency gauge fluctuations are not Landau-damped, and they lead to extra universal scattering and an enhanced universal resistivity at the critical point. Thus the resistivity measured in the usual conditions of $\omega \rightarrow 0$ first, and then $T \rightarrow 0$, will be universally enhanced compared with that calculated in the CFT alone. Quantitative calculations of this effect are hard given that we have very little analytic control over the CFT in the first place (except in a large-$N$ expansion). In any case there will be a universal boson resistivity tensor with $\rho_\Phi$  and  $\rho_{\Phi;xy}$ taking values at least, if not larger than $\frac{h}{e^2}$. 

It follows that as the transition is approached from either side, there will be a universal jump of the resistivity tensor just at the critical point (very similar to that discussed\cite{senthil2008theory,witczak2012universal} for the longitudinal resistivity at the continuous Mott transition). If $\rho_{m; ij} $ is the resistivity tensor in the Fermi liquid side, the measured resistivity as a function of tuning parameter $\delta$ and temperature $T$ will thus satisfy a scaling form analagous to Eqn.~\eqref{eq:condscal}
\begin{equation} 
\rho_{ij} - \rho_{m;ij} = \frac{h}{e^2} {\cal R}_{ij} \left(\frac{\delta}{T^{\frac{1}{\nu z}} } \right) 
\end{equation} 
The  tensor ${\cal R}(x)$ is a universal scaling function of its argument. It has the asymptotic values
\begin{eqnarray} 
{\cal R}_{ij} (x \rightarrow 0) & \rightarrow & r_{ij} \\
{\cal R}_{ij} (x \rightarrow \infty) & \rightarrow & 0 \\
{\cal R}_{ij} (x \rightarrow - \infty) & \rightarrow & 2\epsilon_{ij}
\end{eqnarray} 
where $\epsilon_{ij}$ is the antisymmetric tensor in two dimensions with $\epsilon_{xy} = 1$, and $r_{ij}$ is a universal constant tensor. 
The scaling form implies a crossover temperature $T^*$ for the universal jump which behaves as 
\begin{equation} 
T^* \sim |\delta|^{\nu z} 
\end{equation} 
We have already argued that $z = 1$ and $\nu > 2/3$ for the theory to apply. 


Based on the discussion above, we illustrate the behavior of $\rho_{c;xx}$ and $\rho_{c;xy}$ for the physical electrons at zero temperature in Fig \ref{fig:observable}.  Obviously there is a peak of the longitudinal resistance of order  $\sim \frac{h}{e^2}$. Meanwhile $\rho_{c;xy}$ also has a universal jump of roughly the same order upon approaching the transition from the Fermi liquid. A sketch of the expected $T$-dependence of the resistivities is shown in Fig. \ref{fig:T_rho}.

The universality of the  resistivity jump was considered in Ref. \onlinecite{barkeshli2014continuous} but was suggested to not happen based on the possibility of mixing between longitudinal and transverse responses. We however argue that such mixing will not occur for the resistivities which are measured in the  $q = 0, \omega \rightarrow 0$ limit.  The transport properties are encoded in the polarization tensor $\Pi_{\mu \nu}(\omega,\mathbf q)$ which needs to satisfy the equation $q_\mu \Pi_{\mu \nu}=0$ due to gauge invariance.  Then if we take $q=(\omega, \mathbf q=0)$, we must have $\Pi_{0i}(\omega,\mathbf q=0)=0$ for any $i=1,2$ and non-zero $\omega$. At $\mathbf q=0$, the tensor $\Pi_{\mu\nu}$ is then block diagonal and we can just look at $\sigma_{ij}(\omega)=\frac{1}{i\omega}\Pi_{ij}(\omega,\mathbf q=0)$. The Ioffe-Larkin rule $\Pi^{-1}_c=\Pi^{-1}_b+\Pi^{-1}_f$ then leads to the equation $\rho_c=\rho_\Phi+\rho_f$ for the $2\times 2$ resistivity tensor at any frequency $\omega$ (including the dc limit $\omega \rightarrow 0$), and this leads to the universal resistivity jump. 

Furthermore, as already discussed in Ref. \onlinecite{barkeshli2014continuous}, the electronic compressibility $\kappa = \frac{dn}{d\mu}$ will vanish at the critical point. As a function of $T$, we expect $\kappa \sim T$ right at the quantum critical point. This vanishing critical compressibility at the phase transition between two compressible phases may offer another fruitful near-term experimental signature of this quantum critical point. 

  Finally, the thermal conductivity $\kappa_{xx}=\kappa_{xx,\Phi}+\kappa_{xx,f}$ comes from a summation of two parton sectors. $\kappa_{xx,\Phi}=0$ in both CFL and FL phases, and for the f-Fermi sea, the conductivities obey the Wiedemann-Franz law, i.e. $\kappa_{xx,f}/T=L_0\sigma_{xx,f}$ with $L_0=\pi^2k_B^2/(3e^2)$ the Lorenz number. One arrives at the conclusion that $\kappa_{xx}/T=L_0\sigma_{xx,f}$.   Taking the inverse we find 
\begin{align}
\label{eq:mWF}
  T/\kappa_{xx}=\rho_{xx,f}/L_0=\rho_{xx}/L_0,  
\end{align} 
where we note that $\rho_{xx}=\rho_{xx,f}$ from the Ioffe-Larkin rule,and $\rho_{xx,f}\approx 1/\sigma_{xx,f}$ in both CFL and FL phases. This differs from the standard Wiedemann-Franz law that states $\kappa_{xx}=L_0T \sigma_{xx}$. In particular the CFL phase strongly violates\cite{wang2016half} the standard Wiedemann-Franz relation (as $\sigma_{xx}\approx e^4\rho_{xx}/(4h)^2$ so $\kappa_{xx}\propto 1/\sigma_{xx}$). In the FL phase, as $\sigma_{xx}\approx 1/\rho_{xx}$, the Wiedemann-Franz law as well as eq \eqref{eq:mWF} hold. 

Eq \eqref{eq:mWF} generally does not hold near the critical point where both $\kappa_{xx,\Phi/f}\neq 0$ and there is no simple additive relation between the electron conductivity and conductivities of each parton sector. At the critical points of CFL-CDW$^*$ and CDW$^*$-FL transitions, there will presumably be a universal jump of $\kappa_{xx}/T$, from the boson sector $\Phi$. It would be interesting in the future to investigate the universal behavior of $\kappa_{xx}$ near the critical points of CFL-FL and CFL-CDW$^*$,CDW$^*$-FL transitions, which was similarly discussed in metal-insulator transitions in 3D semiconductors from localization\cite{potter2012quantum}. 

Eq. \eqref{eq:mWF} is also violated in the intermediate CDW$^*$ phase, as there we still have $\kappa_{xx}/T=L\sigma_{xx,f}$ (since $\sigma_{xx,\Phi}=0$ in the Mott insulator for $\Phi$) while the electrical resistivity diverges. We showed the relation schematically in fig \ref{fig:fT_rho}. Thus as expected the pinned CDW$^*$ phase will have metallic thermal transport despite being an electrical insulator.

\subsection{Fluctuating CDW order and SO(3) symmetry}

 From our previous discussion, there is an emergent SO(3) symmetry rotating the three CDW order parameters $(n_1,n_2,n_3)$ with momentum $\mathbf Q=\mathbf M_1, \mathbf M_2, \mathbf M_3$.  At the QCP, we expect power law correlation of these CDW orders:

\begin{equation}
  \langle O^a_{CDW} (\mathbf r) O^b_{CDW} (\mathbf r') \rangle=\delta_{ab} \frac{1}{|\mathbf r-\mathbf r'|^{2\Delta_{CDW}}}
\end{equation}
with $a=1,2,3$. We presented arguments that suggest $1 < \Delta_{CDW} < 2$

The existence of slow CDW fluctuations at these wavevectors is remarkable given that a CDW is absent in both the CFL and in the FL phase\footnote{Similar phenomena are found at other deconfined critical points\cite{senthil2004quantum}}. 
Of course in both phases there are power law correlations associated with $2K_F$ fluctuations of the Fermi surface. The slow critical CDW fluctuations have a completely different origin from these $2K_F$ fluctuations and are unrelated to any specific features of the Fermi surface.  Experimentally one may search for static CDW orders with period $2a$ pinned at impurities at the QCP.  

\subsection{Disorder effects} 
Both the FL and CFL phases will survive for weak disorder.  At the critical point, there will be a linear coupling of the disorder to the slow CDW fluctuations. As discussed earlier this coupling is relevant if $\Delta_{CDW} < 2$ which we have already argued is very likely to be true. Thus we expect that weak disorder is a relevant perturbation at the critical point, and leads to a pinning of the fluctuating CDW order. The bosons $\Phi$ will then enter an insulating phase with zero Hall conductivity that intervenes between the superfluid and the $\nu = 1/2$ FQHE.  In the electronic system, it follows that with weak disorder there also is an  intermediate phase between the FL and CFL. This phase has remarkable properties. It is a pinned CDW insulator where the CDW order parameter is locally present but is randomly  oriented across the sample. Despite this the $f$-fermions will continue to be mobile (ignoring Anderson localization effects which will not set in upto very large length scales for weak disorder\footnote{The random pinning of the CDW will occur on a length scale that grows as a power of the disorder strength while the localization length will be exponentially large in the disorder strength. Hence for weak disorder it is legitimate to ignore localization effects. }).   Thus we have an  intermediate exotic CDW insulator phase with electrically neutral diffusive fermions. Experimentally electrical transport will just see an intermediate insulating phase between FL and CFL; however should thermal transport be possible, a large thermal conductance from the neutral fermions can be detected. We will discuss the possibility of this exotic CDW insulator in the clean limit in the next section.

\section{CFL to CDW* transition at $\nu=-\frac{1}{2}$}
\label{sec:1/2cflcdw}
In this section we return to the clean limit and discuss an exotic CDW insulator phase with a coexisting neutral fermi surface.  This phase falls under the category of CDW$^*$ and we will use that nomenclature  to emphasize that there is a neutral Fermi surface in contrast to a conventional CDW insulator. Note that the name CDW$^*$ encompasses a variety of exotic CDW states with coexisting fractionalized excitations. Thus the CDW$^*$ of this section is distinct from the one encountered in Sec. \ref{subsec:fqahtocdw*}. 
We also note that the phase we discuss was also considered in Ref. \onlinecite{barkeshli2014continuous} (and labelled a Gapless Mott Insulator) though they did not emphasize the presence of CDW order. As we will see the CDW has some important consequences. 

\subsection{CDW*: construction, properties  and transition to CFL}

We continue to use the general framework of the parton construction outlined in Sec. \ref{sec:prtnfrmwrk} by writing $c = \Phi f$ and placing $f$ in a `mean field' Fermi liquid state. If we now put the boson $\Phi$ in a CDW insulating state, we get the gapless CDW$^*$ insulating state with a coexisting neutral fermi surface that is coupled to a dynamical $U(1)$ gauge field. 
A unified framework to discuss this state, the CFL, and the nature of the transition is thus naturally obtained by considering the theory for the evolution between the bosonic $1/2$ FQHE state and the bosonic CDW insulator. This evolution is completely analogous to the one considered in the warm-up Sec. \ref{subsec:warmup}. The method used there yields the following action for the boson sector 
\begin{align}
    S_b&=\int dt d^2x \sum_{I=1,2}|(\partial_\mu -i b_\mu)\Phi_{vI}|^2-\frac{2}{4\pi} b db +\frac{1}{2\pi}(A+a) db \notag \\ 
    &~~~-s \Phi_{vI}^\dagger \Phi_{vI} -g (|\Phi_{v}|^2)^2-\lambda (n_1^2+n_2^2+n_3^2)
    \label{eq:Sb_CFL_CDW}
\end{align}
where $\Phi_v=(\Phi_{v1},\Phi_{v2})^T$, and $n^a = \Phi_v^\dagger \sigma^a \Phi_v$.  The two critical bosons $\Phi_{v1},\Phi_{v2}$ form a two dimensional representation of the projective translation symmetry $T_1 T_2=-T_2 T_1$.    We assume the following transformation rule. $T_1: \Phi_v \rightarrow -i\sigma_1 \Phi_v$ and $T_2: \Phi_v \rightarrow -i \sigma_3 \Phi_v$.  It is then clear   that $(n_1,n_2,n_3)=(\Phi_v^\dagger \sigma_1\Phi_v, \Phi_v^\dagger \sigma_2 \Phi_v, \Phi_v^\dagger \sigma_3 \Phi_v)$ represent three CDW orders at momentum $\mathbf Q=\mathbf M_1, \mathbf M_2, \mathbf M_3$. Under the $C_3$ symmetry,  $\Phi_v \rightarrow e^{-i \frac{\sigma_1+\sigma_2+\sigma_3}{\sqrt{3}}\frac{2\pi}{3} }\Phi_v$ which rotates the three CDW orders to each other.   Note that a general quartic term can be written as $(\Phi^\dagger_v \sigma_\mu \Phi_v)(\Phi^\dagger_v \sigma_\nu \Phi_v)$ with $\mu,\nu=0,1,2,3$. Translation symmetry $T_1$ and $T_2$ guarantees that $\mu=\nu$. The $C_3$ symmetry makes sure that the coefficients for the $\mu=\nu=1,2,3$ terms are the same.

If $s>0$ in eq.~\eqref{eq:Sb_CFL_CDW}, then $\Phi_v$ is gapped and we are left with  the action $S_b=\int dt d^2x -\frac{2}{4\pi} \alpha d \alpha+\frac{1}{2\pi} a d \alpha$, which leads to the standard theory of CFL when added to eq.~\eqref{eq:CFL_FL_Sf}. On the other hand, when  $s<0$,  $\langle \Phi_v \rangle \neq 0$ which higgses the gauge field $b_\mu$.    The $\Phi_v$ condensation leads to a non-zero CDW order parameter $\vec n=\Phi_v^\dagger \vec \sigma \Phi_v$. Meanwhile, now the bosonic sector is trivially gapped and can be ignored.  The result is the low energy effective action:
\begin{align}
S&=\int dt d^2x f^\dagger(t,\mathbf x)(i\partial_t+i a_0+\mu)f(t,\mathbf x) \notag \\ 
&~~~+\frac{\hbar^2}{2m}f^\dagger(t,\mathbf x) (\mathcal{\nabla}+i\mathbf a)^2 f(t,\mathbf x)
\label{eq:exotic_CDW}
\end{align}

  The neutral Fermi surface here is inherited from the CFL phase and these $f$ could roughly still be identified with composite fermions carrying dipole moment. A rough but useful analogy is with the spinon fermi surface state of a frustrated Hubbard model close to the Mott transition. The spinons can be roughly viewed as electrons that have lost their electrical charge through fractionalization. In the present case, the neutral fermions of the CDW$^*$ state can probably be viewed as composite fermions that have lost their attached flux through fractionalization.  

  Because there is a long ranged CDW order, the neutral Fermi surface also gets reconstructed. Just across the phase transition to the CFL (assuming it is second order or weakly first order), we expect  that the neutral Fermi surface remains gapless because the CDW order is still small.  But upon moving deeper into the phase, the neutral Fermi surface gradually disappears and the CDW* phase may evolve to a conventional CDW insulator phase through a Lifshitz transition.

\subsection{Remarks on the CFL-CDW$^*$ phase transition}
The structure of the theory near the phase transition is similar to the CFL-FL transition. However we need to ask if the $f$-fermi surface can dynamically decouple from the critical boson sector. The dangerous couplings are those of the CDW order parameter to the Fermi surface. As argued in Sec. \ref{subsec:warmup}, it seems likely that the scaling dimension $\Delta_{CDW} < 1$ (as it is already $1$ in the mean field, and gauge fluctuations will tend to only decrease it further). 
Then the Landau damping of the CDW fluctuations is relevant, and there is no dynamical decoupling between the $f$ and $\Phi$ sectors.
The fate of the transition is unclear in that case (even if the boson sector by itself is critical). It is possible that the Landau damping eventually drives the transition weakly first order.

In the present of disorder, the CDW$^*$ will evolve into the pinned CDW$^*$ phase with no long range charge order. There will however  be a sharp transition which will be second order. The resistivity tensor will have a universal jump as this critical point for reasons similar(see also the discussion below)  to that at the CFL-FL transition approached from either side, which again can be an experimental fingerprint of the transition.

\subsection{Experimental signatures}
 Over the years neutral Fermi surfaces have been proposed in different contexts and ideas proposed for their experimental detection. Many of these ideas will be relevant for the present CDW* phase as well.  We highlight a few: (I) A metallic thermal conductivity $\kappa_{xx}$.  (II) Possible quantum oscillations\cite{motrunich2006orbital,sodemann2018quantum} close to the critical point. Note that the bosonic sector has a small gap just across the phase transition (if at most weakly first order) and still have a finite diamagnetism, which leads to the locking of the internal flux to the external magnetic field: $b=da=\alpha B$, where $B$ is the external magnetic field. We expect $0<\alpha<1$ and the neutral Fermi surface can show quantum oscillations under the internal flux $b$. (III) Charge Friedel oscillations\cite{mross2011charge} associated with $2K_F$ features of the neutral Fermi surface.

Should the CFL-CDW* transition be only weakly first order, we can understand electrical transport in its vicinity using, as before,  the Ioffe-Larkin rule for the $2\times 2$ resistivity tensor. 

\begin{equation}
    \rho_c=\rho_\Phi+\rho_f
\end{equation}
where $\rho_a=\begin{pmatrix} \rho_{a;xx} & \rho_{a;xy} \\ \rho_{a;yx} & \rho_{a;yy}\end{pmatrix}$.

\begin{figure}
\adjustbox{trim={.22\width} {.05\height} {.1\width} {.48\height},clip}
   {\includegraphics[width=1.\textwidth]{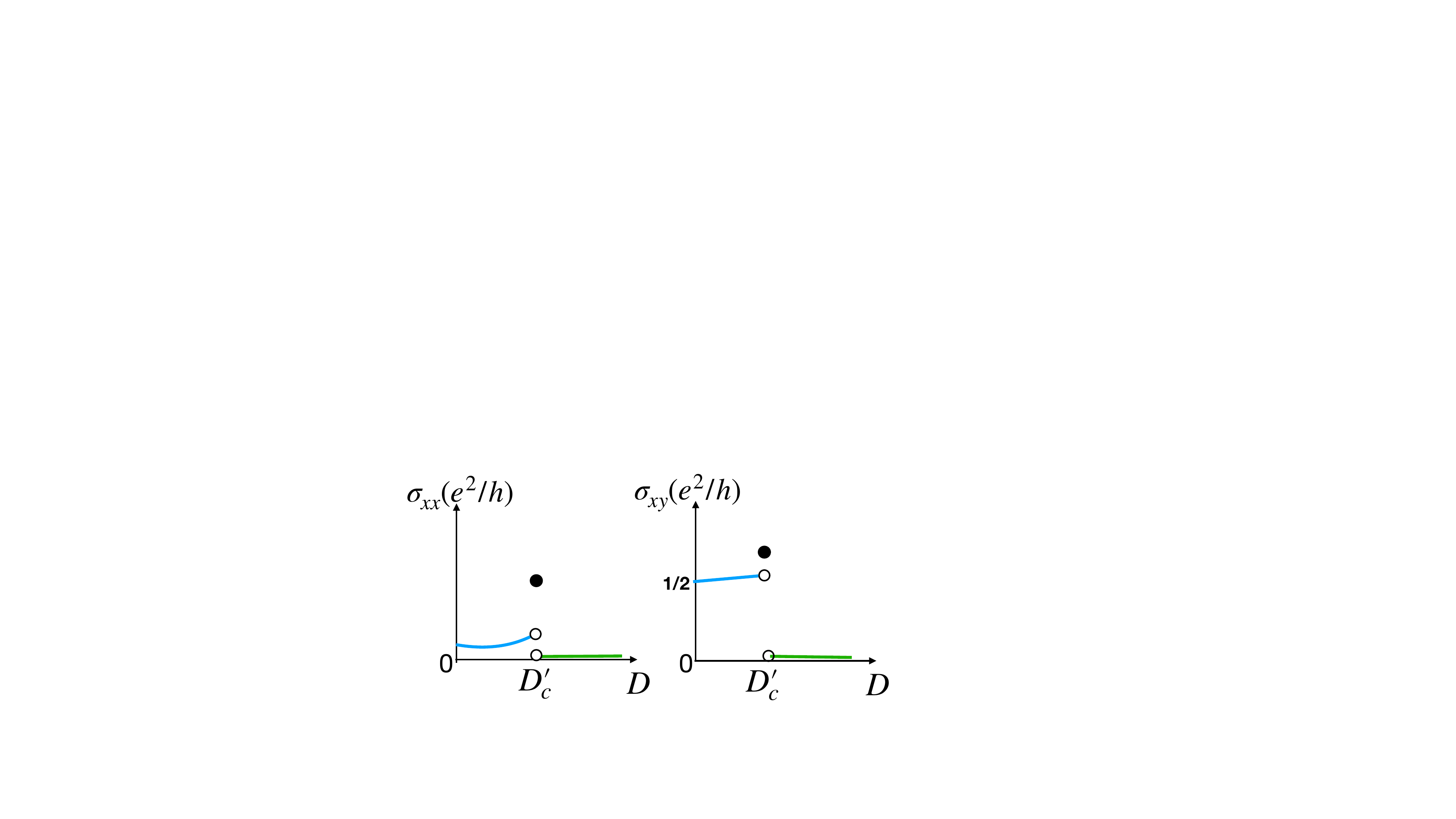}}
    \caption{ Conductivities across the CFL-CDW$^*$(green) transition. $\sigma_{xx,xy}$ both vanish in the CDW$^*$ phase.} 
    \label{fig:cdw_rho}
\end{figure}

$\rho_f$ is again smooth across the transition:

\begin{equation}
    \rho_f=\begin{pmatrix} \rho_f & \rho_{f;xy} \\ -\rho_{f;xy} & \rho_f \end{pmatrix}
\end{equation}
where $\rho_f$ and $\rho_{f;xy}$ are non-universal numbers.  We expect $\rho_f<<\frac{h}{e^2}$ and  $\rho_{f;xy}<<1$  in the weak disorder regime.  In the CFL phase, we have $\rho_\Phi=\begin{pmatrix}0 & -2 \\ 2 & 0 \end{pmatrix} \frac{h}{e^2}$.   But now in the CDW* side, $\rho_\Phi=\begin{pmatrix} +\infty & 0 \\ 0 &+\infty \end{pmatrix}$ because the bosonic sector is in a trivial CDW insulator.  Exactly at the critical point, the CFT nature of the $S_b$ leads to a universal tensor $\rho_\Phi=\frac{1}{\sigma_b^2+\sigma_{b;xy}^2} \begin{pmatrix} \sigma_b & -\sigma_{b;xy} \\ \sigma_{b;xy} & \sigma_b \end{pmatrix}$ where $\sigma_b$ and $\sigma_{b;xy}$ are universal numbers at order of $\frac{e^2}{h}$.   We illustrate the behavior of $\sigma_{c;xx}$ and $\sigma_{c;xy}$ for physical electrons at zero temperature in Fig. \ref{fig:cdw_rho}.   $\rho_{xx}$ goes through a metal-insulator transition with a universal jump.

 Finally we remark that one can contemplate more exotic scenarios where, despite the presence of the Landau damping of the CDW fluctuations, the transition stays continuous but without dynamical decoupling between the $f$-fermi surface and the $\Phi$-bosons. Given our lack of control of the theory in this scenario, we will not speculate on it further in this paper.

 We have discussed CFL to FL and CFL to CDW* transition.  In Appendix.~\ref{append:FL-CDW*_tri_critical}, we provide a critical theory for the FL to CDW* transition and a tri-critical theory that captures the coming together of all three phases  (CFL, FL and CDW*).

\section{CFL to FL transition at $\nu=-\frac{3}{4}$ filling}

\label{sec:3/4cflfl}
Numerical work\cite{dong2023composite,goldman2023zero} on models of the tMoTe$_2$ system find evidence for a CFL phase at $\nu = -\frac{3}{4}$ in a range of displacement field, and a FL outside of that range. Here we discuss the phase transition between these two phases. A theory for a direct CFL-FL transition at this filling has not appeared previously in the literature. 

The  $\nu=-\frac{3}{4}$ state can be reached by starting in the from a $\nu=-1$ IAQH state, and doping electrons to  a filling $\nu_e =\frac{1}{4}$. Thus the CFL at this filling can be obtained as the particle-hole conjugate of one at filling $\nu_e =\frac{1}{4}$. Accordingly we will focus on the CFL-FL transition at $\nu_e = \frac{1}{4}$. 

Using the usual parton construction  $c=\Phi f$, the CFL and FL states correspond to a Laughlin state and superfluid state for the $\Phi$, respectively. The Laughlin state is described by a  Chern-Simons Lagrangian of  a $U(1)$ gauge field at level $4$, i.e. 
\begin{equation}
 {\cal L} =    -\frac{4}{4\pi}\alpha d\alpha+\frac{1}{2\pi}\alpha dA
\end{equation}
 The full action for the electrons again consists of $S=S_b+S_f+S_{bf}$ where the $f$-fermions form a Fermi surface at filling $\nu=1/4$. 
\begin{align}
\label{eq:3/4_f}
    S_f=\int dt d^2x f^\dagger (t,\mathbf x)(i\partial_t+ia_0+\mu)f(t,\mathbf x)\nonumber\\+\frac{\hbar^2}{2m}f^\dagger(t,\mathbf x)(\mathbf \nabla+i \mathbf a)^2 f(t,\mathbf x),
\end{align}
 
The question of the CFL-FL transition then crucially relies  on the nature of the corresponding transition in the $\Phi$ sector. Here we show that a direct, second-order $1/4$ Laughlin-superfluid transition for a boson system is possible.  We begin with a  parton representation of $\Phi$ in terms of four fermions
\begin{equation} 
\Phi=f_1 f_2 f_3 f_4, 
\end{equation} 
This representation introduces an $SU(4)$ gauge redundancy but we will assume that the mean field states of interest break it to $U(1) \times U(1) \times U(1)$. Thus we will have 3 dynamical $U(1)$ gauge fields appearing in the theory.  
At boson lattice filling  $\nu_\Phi=1/4$, the gauge constraints fix the filling of each of the partons $f_i$  to also be   $\nu_{fi}=1/4 (i=1\cdots 4)$. We assign the charge of the background gauge field $A_b$ to $f_1$. (In the eventual application to the electronic transition, $A_b$ will be replaced by $a +A$).  When each $f_i$ fills a Chern band with $C=1$, the resulting state is the desired $1/4$ Laughlin state as shown below. When two of the $f_i$ fill a Chern band with $C=-1$ and the other $2$ remain in the $C=1$ band, the resulting state describes a superfluid phase for $\Phi$. The key is to construct a mean-field state where the $f_i$ partons can  completely fill a  band at filling $1/4$ and, simultaneously, to allow for a  a change of Chern number from $1$ to $-1$ for two of the $f_i$-partons. We achieve this by making the $f_i$ to move in a background flux such that  the unit cell is quadrupled, while still preserving the full translation symmetry for $\Phi$.

We take the mean field for the $f_i$  to feel flux $\phi_{\pm}=\pi/2\pm \theta_d$ for alternating plaquettes (parallellogram) on the triangular lattice. Within one plaquette, the flux around upward/downward triangles is distributed as $\phi_{\pm}+\theta,-\theta$, respectively. The unit cell is quadrupled and thus at $\nu_{fi} = \frac{1}{4}$, the partons could entirely fill one band. We choose a mean field such that the $f_{1,2}$ each fill a Chern band with $C=1$ with $\pi/2$ flux per plaquette, $\theta_d=0$. The $f_{3,4}$ feel flux $\phi_{\pm}$, respectively, around a single plaquette, such that combined they see $\phi_++\phi_-=\pi$ flux per plaquette. With this choice, one can tune the $\phi_{\pm}$ so $f_{3,4}$ realizes a transition from $C=1$ to $C=-1$ through a band closing with $2$ Dirac cones. The detailed mean-field is listed in Appendix \ref{app:mf_14}. Crucially the action of one of the translation symmetries exchanges $f_{3,4}$ so that their Chern numbers change simultaneously. Thus this construction gives (at the mean field level) a realization of the  desired second order transition from Laughlin state at $\nu=1/4$.

\subsection{Critical theory}

Let us show that a field theory  derived from this parton construction indeed describes both the Laughlin and superfluid states at $1/4$ filling,  and their phase transition.  The $\Phi=f_1 f_2 f_3 f_4$ construction introduces three $U(1)$ gauge fields $a_1,a_2,a_3$, together with the background $A_b$ field which we assign to couple to $f_1$. Thus $f_1$ couples\footnote{It may be useful to note that $A_b, a_2$ are ordinary $U(1)$ gauge fields while $a_1, a_3$ are spin$_c$ connections. 
} to   $A_b+a_1$, $f_2$ to $-a_1+a_2$, $f_3$ to $-a_2+a_3$, and $f_4$ to $-a_3$. 
The parton mean field described above then leads to the effective Lagrangian 
\begin{equation}
    \mathcal L=\sum_{a=1,2,3,4}\mathcal L_a
\end{equation}
where
\begin{equation}
    \mathcal L_1=-\frac{1}{4\pi}\alpha_1 d \alpha_1+\frac{1}{2\pi} \alpha_1 d \left(a+a_1+A_b \right)
\end{equation}

\begin{equation}
    \mathcal L_2=-\frac{1}{4\pi}\alpha_2 d \alpha_2+\frac{1}{2\pi} \alpha_2 d (a_2-a_1)
\end{equation}

\begin{align}
    \mathcal L_3&=\bar \psi_1\left(-i\gamma_\mu (\partial_\mu \sigma_0-b_{1;\mu} \sigma_0 )\right)\psi+m \bar \psi_1 \psi_1 \notag \\ 
    &~~~- \frac{1}{4\pi} b_1 db_1 +\frac{1}{2\pi} b_1 d \alpha_3+\frac{1}{2\pi} \alpha_3d (a_3-a_2)
\end{align}

\begin{align}
    \mathcal L_4&=\bar{\psi_2}\left(-i\gamma_\mu (\partial_\mu\sigma_0- b_{2;\mu} \sigma_0) \right) \psi_2+m \bar{ \psi_2}  \psi_2 \notag \\ 
    &~~~- \frac{1}{4\pi} b_2 db_2 +\frac{1}{2\pi}  b_2 d \alpha_4+\frac{1}{2\pi} \alpha_4d (-a_3)
\end{align}

Integrating $a_1,a_2,a_3$ locks $\alpha_1=\alpha_2=\alpha_3=\alpha_4=\alpha$.  Thus the final Lagrangian is
\begin{align}
    \mathcal L_{critical}&=\sum_{i=1,2}\bar \psi_i(-i\gamma_\mu (\partial_\mu \sigma_0-b_{i;\mu} \sigma_0))\psi_i+m \sum_i \bar \psi_i \psi_i \notag \\
    &~~~-\frac{2}{4\pi} \alpha d \alpha -\frac{1}{4\pi}(b_1db_1+b_2db_2) \notag \\
    & +\frac{1}{2\pi} (A_b +b_1+b_2)d\alpha 
\end{align}
where $\psi_1$ and $\psi_2$ both have two flavors, so that in total there are four Dirac cones. When $m>0$, integrating out $\psi_i$ gives $\frac{2}{4\pi}(b_1db_1+b_2db_2)$. Further integrating out $b_{1,2}$ results in the Laughlin state described by 
\begin{align}
    \mathcal L=\frac{-4}{4\pi} \alpha d\alpha+\frac{1}{2\pi} A_b d\alpha.
\end{align}
When $m<0$, integrating out $\psi_{1,2}$ produces no new Chern-Simons terms.  Further integrating out $b_{1,2}$ gives a Chern-Simons term for $\alpha$ that cancels the term from $f_{1,2}$ Chern bands,
\begin{align}
    \mathcal L=\frac{1}{2\pi} A_b d\alpha,
\end{align}
which implies that the $\Phi$ forms a superfluid.

We have the following gauge  invariant operators: $\vec n_1= \bar \psi_1 \vec \sigma \psi_1$, $\vec n_2=\bar \psi_2 \vec\sigma \psi_2$.  We can also organize them into $\vec n_+=\vec n_1+\vec n_2$ and $\vec n_-=\vec n_1-\vec n_2$. There is an emergent $SO(3)\times SO(3) \times Z_2$ internal symmetry. Each $SO(3)$ factor corresponds to flavor rotations of one of the two massless Dirac sectors. The $Z_2$ interchanges the two Dirac sectors. This symmetry originates in the particular realization of lattice translations on the partons. 

The detailed translation action depend on the mean-field construction.  In general rotation symmetry is broken, resulting in two nonequivalent translation directions along $T_{1,2}$. We list here one possible translation actions on the Dirac fermions and the gauge invariant fermion bilinears, $\vec n_1= \bar \psi_1 \vec \sigma \psi_1$, $\vec n_2=\bar \psi_2 \vec\sigma \psi_2$:
\begin{align}
    T_1:\psi_i\rightarrow i\sigma_z \psi_i (i=1,2)
    \nonumber\\
    (n_i^x,n_i^y,n_i^z)\rightarrow (-n_i^x,-n_i^y,n_i^z)\nonumber\\
    T_2:\psi_3\rightarrow -\psi_4,\psi_4\rightarrow \sigma_x \psi_3 \nonumber\\
    \vec n_1\rightarrow \vec n_2,(n_2^x,n_2^y,n_2^z)\rightarrow (n_1^x,-n_1^y,-n_1^z).
\end{align}
The translations obey the projective relations $T_1T_2^2 T_1^{-1}T_2^{-2}=-1$ on the Dirac fermions. For the gauge invariant bilinears, one could organize them according to their momenta as:(these are measured in the reciprocal lattice vector associated with translations in fig \ref{fig:triangle}) 
\begin{align}
    (\pi,0): n_1^x+ n_2^x,
    (\pi,\pi):n_1^x- n_2^x,\nonumber\\
    (\pi,-\pi/2): n_1^y+i n_2^y,
    (\pi,\pi/2): n_1^y-i n_2^y,\nonumber\\
    (0,-\pi/2):n_1^z+i n_2^z,(0,\pi/2):n_1^z-i n_2^z.
\end{align}
Note that the critical theory explicitly breaks $C_3$ symmetry. It is  not ruled out though that there may be a second order transition that accommodates both translations and $C_3$.

 There is also a gauge invariant monopole operator $\mathcal M=\mathcal M_{\alpha} \mathcal M_{b_1} \mathcal M_{b_2}$.   Note that $\mathcal M_{\alpha}^\dagger $ carries $q_{\alpha}=2, q_{b_1}=1, q_{b_2}=1$.  Bare operator $\mathcal M^\dagger_{b_1}$ carries $q_{\alpha}=1, q_{b_1}=-1$ because one needs to combine a zero mode $\psi_1^\dagger$ to make it neutral under $b_1$.  Similarly $\mathcal M_{b_1}^\dagger$ carries $q_{\alpha}=1$ and $q_{b_2}=-1$.  Together $\mathcal M^\dagger$ is gauge invariant and represents the boson operator $\Phi$ with charge $1$ under $A_b$.
$\mathcal M^\dagger_{\alpha} \mathcal M_{b_I}^\dagger \mathcal M_{b_I}^\dagger \bar \psi_{I;\sigma} \psi_{3-I;\sigma'}$ for $I=1,2$ are also gauge invariant operators, which carry unit physical charge and are of higher scaling dimension than $\mathcal M$. 

The field theory above thus provides for a route to a second order transition between Laughlin and superfluid states of bosons at $1/4$ filling. Now we can just follow the same steps as at $1/2$ filling to discuss the CFL-FL transition. As the logic is identical to what we already described at $1/2$ filling we will not repeat them here. The exponent bounds $\bar{\nu} > \frac{2}{3}$, and $\Delta_{n_i} > 1$ are once again required for the $f$ and $\Phi$ sectors to decouple.



 The various physical properties discussed for the CFL-FL transition at $1/2$ filling will also hold for the transition at $3/4$ filling with only some obvious modifications.  First  there is a $\nu=-1$ background integer quantum anomalous hall(IQAH) state for the holes which must be incorporated in thinking about the net transport. Second, in using the Ioffe-larkin rule, the Hall resistance of the boson in the CFL is $\frac{4h}{e^2}$. Apart from these minor modifications, we will get the phenomena of the universal resistivity tensor jump, a vanishing compressibility, and the appearance of an intermediate  pinned CDW$^*$ with weak disorder. 
 


\subsection{A different intermediate composite fermi liquid} 
In this subsection we briefly describe an alternate route for the evolution between the CFL and FL at $\nu_e = \frac{1}{4}$. The CFL described thus far in this section is the usual one that in the Landau level setting is obtained by binding four vortices to the electron. Let us denote this state CFL$_4$. If these 4  vortices all completely unbind from the electron we get the FL. But we can ask if there is an intermediate translation invariant phase where only 2 vortices bind to each electron to form distinct composite fermions. We will denote the resulting composite fermi liquid CFL$_2$. Remarkably we will see that a CFL$_2$ state can exist at $1/4$ filling and be translation invariant. 

In terms of the $c = \Phi f$ construction, we seek a phase where the $\Phi$ form  a topologically ordered state described by $U(1)_2$ Chern-Simons theory and the $f$ in a Fermi surface state. Now for ordinary bosons (with linearly realized lattice translations) the $U(1)_2$ theory cannot be realized as a translation invariant state at $1/4$ filling as it violates filling constraints.  However as the $\Phi$ couple to the dynamical gauge field $a$, there is a possibility of a mean field state where $\Phi$ and $f$ both see $\pi$ flux through each plaquette. Then a $U(1)_2$ topological order for $\Phi$ does not violate translation symmetry. The key point is that due to the background $\pi$ flux, the unit cell for $\Phi$ is doubled. The filling of this doubled unit cell is $1/2$, and a $U(1)_2$ state is allowed.   But of course translation symmetry action on the physical electrons is not broken, and all gauge-invariant observables will be translation invariant.  

An alternate  simple way to see that CFL$_2$ can exist at $1/4$ filling is as follows: consider a Landau level in a periodic potential $V$  with one flux quantum per unit cell. At $V =0$ (no periodic potential), at half filling we have the usual CFL$_2$ and at quarter filling we have CFL$_4$. 

Now consider starting at $1/2$ filling and moving away to lower filling. The composite fermions (denote CF$_2$) of CFL$_2$ now see an effective field 
\begin{equation} 
B^* = B(1-2\nu) 
\end{equation} 

At $V$ non-zero, and at $1/4$ filling, the composite fermion (CF$_2$) flux per unit cell is $\pi$. 
Now these CF$_2$ fermions can form bands again and form a CFL$_2$ state. 

The quarter-filled CFL$_2$ state will have a Hall resistance close to $\frac{2h}{e^2}$ unlike the CFL$_4$ state (whose Hall resistance will be close to $\frac{4h}{e^2}$). Further 
generically the Fermi surface of this CFL$_2$ quarter-filled state will be very different from the CFL$_4$ state, leading to subtle differences in $2K_f$ structures but to cleanly tell the difference may be hard. 

In any case this state can exist and may represent a good intermediate state between CFL$_4$ and the FL. Essentially the electron goes from binding 4 vortices to binding 2 vortices to binding no vortices. 

\section{Discussion}

Here we reiterate some of our results and comment on some directions for future research. 
We have provided theories for various possible continuous phase transitions out of, and phases proximate to,   familiar quantum Hall phases in the twisted MoTe$_2$ bilayer.  At zero displacement field $D$, we have the usual quantum Hall phases as in the familiar lowest Landau level.  When increasing $D$, the bandwidth of the Chern band increases and the quantum phases transit to more conventional phases without Hall conductivity.

For incompressible FQAH phases, we obtained quantum Ginzburg-Landau theories that describe the competition with CDW phases, and are suitable for addressing universal aspects of the phase transition. At filling $\nu=-\frac{2}{3}$, a continuous transition between a FQAH phase and a trivial insulator may already have been observed\cite{xu2023observation}. We described two transitions between the FQAH phase and insulating CDW phases, one (CDW-IAQH) where the CDW also has an Integer Quantum Anomalous Hall effect, and another (CDW$^*$) where there is no electrical Hall effect but there is a `dark' topological order of neutral quasiparticles. This neutral topological order has a quantized thermal hall conductivity.  
Such a CDW* phase is indistinguishable from  a trivial insulator in electrical transport. The possibility of such an unconventional CDW phase proximate to the FQAH is interesting to study in future experiments.  Whether we can have a continuous transition directly from the FQAH phase to a conventional CDW phase is not clear to us, especially with strong disorder. 

At filling $\nu=-\frac{1}{2}$, we expect a continuous CFL to FL transition in the clean limit. The transition is protected by a projective translation symmetry. Associated with this,  there must be fluctuating CDW orders at the CFL-FL critical point. We described some experimental signatures, including a universal jump of the resistivity tensor at the critical point, which may be accessible in the near term.  With disorder, we argue that the CDW orders will be pinned. The critical point then gets replaced with an intermediate `pinned CDW$^*$' insulator phase with mobile diffusive neutral fermions.  This phase has a parent in the clean system: an insulating  CDW$^*$ phase with a neutral fermi surface. Theoretically, it will be interesting to explore the possible occurence of this phase in microscopic calculations of model Hamiltonians for the tMoTe$_2$ system.  In current  experiments there is indeed an intermediate insulating phase\cite{park2023observation} between the CFL and the FL phase. However, we need a sample with reduced disorder strength to distinguish between the following two scenarios: (1) A CFL-FL critical point in the clean limit that is covered by a CDW* phase due to disorder  (2) a CFL-CDW*-FL sequence of phases even in  the clean limit.  

In all of our theories, we assumed full valley polarization and only considered a purely spinless model. Full valley polarization is indeed expected at zero displacement field, but valley polarization should get suppressed at larger displacement field.  We have assumed that the loss of valley polarization happens after the critical point we discussed. If the full valley polarization is destroyed before the quantum Hall phase transition, the critical theories will be more complicated.  We note that quantum Hall phases can in principle be compatible with partial valley polarization with inter-valley-coherence (IVC) order. Therefore the possibility of phase transitions at partial valley polarization needs to be checked by future experiments.

There are other theoretical challenges related to phase transitions out of FQAH phases that we have not attempted to attack. A fascinating and long standing question is to ask if a direct continuous transition between an FQAH state and the FL is possible at fixed filling in a translation invariant system. This transition involves the continuous death of the Fermi surface (without changing its area) that is concomitant with the onset of topological order. Such a transition may have been seen in experiments\cite{cai2023signatures,zeng2023integer,park2023observation} on tMoTe$_2$, and is hinted at by numerical studies\cite{reddy2023fractional} as well. A different challenge is to develop a microscopic analytic theory for the evolution out of the quantum Hall phases. We have focused in this paper on low energy effective field theory descriptions which are only suitable for universal physics near the transition. Developing microscopic theories will have to confront another long standing challenge; that of describing many body physics in a single partially filled topological band. For some limited progress in the Landau level context, see Refs. \onlinecite{Read_1998,dong2020noncommutative,dong2022evolution,goldman2022lowest,ma2022quantitative,govcanin2021microscopic}. 
The importance of these old theoretical questions to modern experiments will, we hope, lead to progress toward their solution.

\section{Conclusion}
Quantum Hall phases are arguably the most exotic experimentally established quantum phases of matter. Their presence in the highly tunable moir\'e platform motivates study of proximate phases and associated phase transitions. Perhaps not surprisingly, these phase transitions (and some of the proximate phases) are exotic as well, and indeed realize novel `beyond Landau' criticality that have been explored theoretically for many years.  Thus these moir\'e platforms provide a great experimental opportunity to study these unconventional phase transitions. Further some of the proximate phases inherit a part of the physics of the quantum Hall phases but without the experimental signature of the quantized Hall transport. This may enable experimental realization of such phases in the moir\'e platform which is another long standing goal in condensed matter physics. 

{\em Note added} Since the initial submission of the manuscript, fractional quantum anomalous Hall states and anomalous composite fermi liquids have been reported in a pentalayer graphene moir\'e system. The data on the CFL-FL transition, while preliminary, is consistent with our prediction of a universal peak in the longitudinal resistivity in the clean limit.

\section{Acknowledgement} We thank Eric Anderson, Jiaqi Cai, Heonjoon Park and Xiaodong Xu for discussions on their experimental data. YHZ thanks Yihang Zeng for discussions.  YHZ was supported by the National Science Foundation under Grant No.DMR-2237031. The work by YHZ was performed in part at Aspen Center for Physics, which is supported by National Science Foundation grant PHY-2210452. TS was supported by NSF grant DMR-2206305, and partially through a Simons Investigator Award from the Simons Foundation. XYS was supported by the Gordon and Betty Moore Foundation EPiQS Initiative through Grant No.~GBMF8684 at the Massachusetts Institute of Technology. This work was also partly supported by the Simons Collaboration on Ultra-Quantum Matter, which is a grant from the Simons Foundation (Grant No. 651446, T.S.).

%

\appendix 

\onecolumngrid

\section{Dual theories of the bosonic superfluid-Laughlin transition at half-filling}
\label{app:dual}

As we have discussed in the main text, a crucial building block of the CFL to FL transition at half-filling is the corresponding Laughlin-superfluid transition of the boson $\Phi$.   There are actually several equivalent critical theories for this bosonic phase transition, as was discussed in the context of chiral spin liquid to superconductor transition (expected to be in the same universality class) by two of us\cite{song2022deconfined}. In the following we list these three theories which are dual to each other. For the bosonic Laughlin state at half filling per unit cell, the semion must obey a projective translation symmetry $T_1 T_2=-T_2 T_1$. It turns out that this algebra is enough to constrain the possible critical theory and the symmetry properties of the operators. We also assume $C_3$ rotation symmetry in our discussion, but will argue that $C_3$ is not essential.   We introduce the usual  probe field $A_b$ for the bosonic sector but it will be promoted to be the gauge field $a + A$ in the application to the CFL-FL  transition.

\textbf{SU(2)$_1$ with $2\Phi$} We start from a  theory with an $SU(2)$  gauge field. It is known that the bosonic FQHE phase at $1/2$ filling can be described by a SU(2) gauge theory with a level-1 Chern-Simons term. Then a natural way to get out of it is to Higgs the SU(2) gauge field through condensing a bosonic field $\Phi_i$ ($i = 1,2$) in the fundamental representation. When $\Phi_i$ is gapped, we must recover the familiar $SU(2)_1$ theory: the gapped $\Phi_i$ will evolve into the semion in the FQHE phase. But we know that the semion must satisfy the projective translation symmetry $T_1 T_2=-T_2 T_1$.  This is the reason for the $i$ index to take 2 values (so that $\Phi_1,\Phi_2$ can  satisfy the translation algebra\footnote{The projective translation symmetry $T_1 T_2=-T_2 T_1$ does not have one dimensional representation.}). We choose the basis so that the  projective translation symmetry acts as $-i\sigma_1$ and $-i\sigma_3$ in the $\Phi=(\Phi_1,\Phi_2)$ flavor space: $T_1: \Phi \rightarrow  -i\sigma_x \Phi$, $T_2: \Phi \rightarrow -i \sigma_z \Phi_2$. Here $\sigma_1, \sigma_2, \sigma_3$ are Pauli matrices. The action of $T_1$ and $T_2$ are fixed up to a U(1) factor $e^{i\theta_{1,2}}$ which we will discuss later.  We also know that the semion carries $1/2$ charge under $A_b$, so the critical theory must be have the form:

\begin{align}
    \mathcal L_{SU(2)}&=\sum_{i=1,2}|(\partial_\mu-i\alpha^s_\mu \tau_s-i \frac{1}{2} A_{b;\mu} \tau_0 )\Phi_i|^2-r |\Phi|^2 -\frac{1}{4\pi} Tr[ \alpha \wedge d \alpha+\frac{2}{3} i  \alpha \wedge  \alpha \wedge  \alpha]+\frac{1}{8\pi}A_b d A_b\notag \\ 
    &~~~- \big(g|\Phi^\dagger \Phi|^2+\lambda_0 \mathbf n \cdot \mathbf n-\lambda (n_4^2+n_5^2))\big)
    \label{eq:SU2_theory}
\end{align}

where $\alpha^s_\mu$, $s=0,1,2$ is a SU(2) gauge field.  $\tau_s$ is the Pauli matrix in the color space rotating $(\Phi_{a;1},\Phi_{a;2})$ with $a=1,2$ as the flavor index.  $\sigma_a$ is Pauli matrix in the flavor space rotating $\Phi=(\Phi_1,\Phi_2)$.  $|\Phi|^2=\Phi^\dagger \tau_0 \sigma_0 \Phi$.  $\frac{1}{4\pi} Tr[ a \wedge da+\frac{2}{3}i a \wedge a \wedge a]$ is the Chern-Simons term for $SU(2)$ gauge field with $\alpha=\sum_s \alpha^s \tau_s$. When $r>0$, we have $\Phi_1,\Phi_2$ gapped and we recover the $SU(2)_1$ description of the FQHE phase with the correct response $\frac{1}{8\pi}A_b d A_b$. Because of the projective translation symmetry,  $\Phi_1,\Phi_2$ have the same action and mass terms.

 There are five gauge invariant bilinear operators $\mathbf n=(n_1,n_2,n_3,n_4,n_5)=(\Phi^\dagger \sigma_1 \Phi, \Phi^\dagger \sigma_2 \Phi, \Phi^\dagger \sigma_3 \Phi,\text{Re} \Phi^T \sigma_2 \tau_2 \Phi, \text{Im} \Phi^T \sigma_2 \tau_2 \Phi)$ in addition to $|\Phi|^2$ in this theory.   $n_4+in_5$ apparently carries charge $1$ under $A_b$ and can be identified as the superfluid order parameter. The transformations of $(n_1,n_2,n_3)$ under $T_1,T_2$ are consistent with three CDW order parameters with momentum $\mathbf Q=\mathbf M_1, \mathbf M_2, \mathbf M_3$.  So obviously $C_3$ rotates $(n_1,n_2,n_3)$ to each other.  For the superfluid order parameter $n_4+i n_5$, its action under $T_1$ and $T_2$ depends on the $U(1)$ factor in the action of $T_1, T_2$ on $\Phi$.

 Here we try to constrain the U(1) factor associated with $T_1$ and $T_2$. Let us assume the following transformation rule. $T_1: \Phi \rightarrow -e^{i\theta_1/2} i\sigma_1 \Phi$ and $T_2: \Phi \rightarrow -e^{i\theta_2/2} i \sigma_3$. $C_3$ must act like $\Phi \rightarrow e^{i \theta_3} e^{-i \frac{\sigma_1+\sigma_2+\sigma_3}{\sqrt{3}}\frac{2\pi}{3}} \Phi$ to rotate $(n_1,n_2,n_3)$ to each other. Then the algebra $C_3T_1 C_3^{-1}=T_2$ will fix $\theta_1=\theta_2=\theta$.    The algebra $C_3^{-1}T_1 C_3=T_1^{-1}T_2^{-1}$ further fixes $\theta=0,\pm \frac{2\pi}{3}$.  With $\theta=0,\pm \frac{2\pi}{3}$, one can easily check that the superfluid order parameter $n_4+in_6=\Phi^T \sigma_2 \tau_2 \Phi$ is at momentum $\mathbf Q=\Gamma, K, K'$ in the MBZ. There is actually a quite simple argument to constrain the momentum of the SF order.  Our critical theory only has one gauge invariant order parameter $n_4+in_5$ which can be identified as a SF order parameter.  If there is a C$_3$ symmetry, SF order must form a one dimensional representation of $C_3$, so its momentum can only be at the $C_3$ invariant point: $\Gamma, K, K'$.  If there is a $C_6$ symmetry, the above argument will constrain the SF order parameter at momentum $\Gamma$ because it's the only $C_6$ invariant momentum.  However, there is only $C_3$ symmetry in twisted MoTe$_2$ system, so we are left with three different choices for the SF momentum. However, in this case these three momentums are equivalent as one can freely choose the definition of  the Gamma point. In twisted MoTe$_2$, the band minimum is at either $K$ or $K'$ of the MBZ depending on the valley.  We can choose the choice of $\theta_1=\theta_2=\theta$ in our critical theory to make sure that the SF order parameter is at the appropriate momentum.

 In the last line of eq.~\eqref{eq:SU2_theory} we include the generic quartic terms obeying $C_3$ rotation symmetry.  As $C_3$ rotates $(n_1,n_2,n_3)$ to each other and forbids terms like $-\lambda' (n_1^2+n_2^2)$. The above action also has a $U(1)\times SO(3)$ symmetry with the three CDW orders rotating under an emergent SO(3) symmetry and the U(1) corresponds to the superfluid order. The FQHE-SF transition corresponds to the fixed point with $\lambda>0$. Later we will argue that the C$_3$ breaking term is irrelevant at this fixed point.   We will also discuss another fixed point at $\lambda<0$ and a tri-critical point in the next section when we discuss FQHE to CDW transition.

\textbf{U(1)$_2$ with $2\varphi$} We know that the FQHE phase can also be captured by a U(1)$_2$ theory. Hence we also expect a critical theory with a single U(1) gauge field.  An easy way to get the U(1) theory is to add a term $- \Phi^\dagger \sigma_3 \tau_3 \Phi$ to eq.~\eqref{eq:SU2_theory}.  Then at low energy we only need to keep two critical field $\varphi_1=\Phi_{1;1}$ and $\varphi_2=\Phi_{2;2}^*$. The critical theory reduces to:

\begin{align}
\mathcal L&=|(\partial_\mu -i \alpha_{\mu}- i \frac{1}{2} A_{b;\mu} \sigma_3 )\varphi|^2-r |\varphi|^2 -\frac{2}{4\pi}\alpha d\alpha -g(|\varphi|^2)^2+\lambda |\varphi_1|^2|\varphi_2|^2+ \frac{1}{8\pi} A_b d A_b
\label{eq:u1_theory}
\end{align}
where $|\varphi|^2=|\varphi_1|^2+|\varphi_2|^2$ and Pauli matrix $\vec{\sigma}$ acts in the $\varphi=(\varphi_1,\varphi_2)^T$ space. One can check that $r>0$ and $r<0$ give the FQHE and the superfluid phase (provided that $\lambda>0$ and the condensation $\langle \varphi \rangle \propto (1,1)^T$ in the Higgsed side). Because the two phases are the same as in eq.~\eqref{eq:SU2_theory}, we expect that the U(1) critical theory here is dual to the SU(2) theory in eq.~\eqref{eq:SU2_theory}.  Again $T_1 T_2=-T_2 T_1$ is important to guarantee that there are two critical bosons. $\varphi^\dagger \sigma_3\varphi$ can be identified as the CDW order $n_3$ while $(n_1,n_2)$ are the monopole operators now\cite{song2022deconfined}.  The emergent SO(3) symmetry is not explicit in this formulation.

\textbf{U(1)$_{-1}$ with $2\psi$}  Finally using the boson-fermion duality, from eq.~\eqref{eq:u1_theory}, we can derive the critical theory in terms of Dirac fermions coupled to a $U(1)$ gauge field discussed in Eqn.~\eqref{eq:U1_2psi} of the main text. 
In the Dirac theory, it is natural to expect that the quartic terms are irrelevant so that there is an emergent SO(3) symmetry rotating the three CDW orders $(n_1,n_2,n_3)=(\bar \psi \sigma_1 \psi,\bar \psi \sigma_2 \psi, \bar \psi \sigma_3 \psi)$. This conclusion does not rely on the microscopic $C_3$ rotation symmetry, implying that the $C_3$ breaking term is irrelevant at this fixed point with emergent SO(3) symmetry.

In summary, we provide three equivalent formulations of FQHE to SF transition for physical boson at filling $\nu=\frac{1}{2}$ of a Chern band.  The critical theory is protected by a projective translation symmetry $T_1 T_2=-T_2 T_1$. We assume a $C_3$ symmetry in our discussion, but we argue that small $C_3$ breaking is irrelevant. The critical theory has fluctuation CDW orders at momentum $\mathbf M_1, \mathbf M_2, \mathbf M_3$ with an emergent SO(3) symmetry.  The existence of CDW order at the QCP is remarkable given that it is absent and unrelated to both the FQHE and SF phases in the two sides.  

These three formulations are equivalent in the sense that they are expected to flow to the same fixed point in the IR. But they will not provide equivalent approximate mean field treatments of the transition. For this purpose we expect that the Dirac fermion formulation may be the most suited. The

Then we can obtain $S_b$ by replacing $A_b$ with the field $a+A$ to reach three equivalent descriptions of the CFL-FL transition at half filling for  electrons. The $SU(2)$ gauge theory of course will ahve the strong fluctuations of the non-abelian gauge field compared to the Dirac theory with its $U(1)$ gauge field. The bosonic theory (U(1)$_2$ with $2\varphi$) might be expected to have weaker fluctuations but it does not have manifest $SO(3)$ global symmetry which we have argued is a feature of the true IR fixed point. Thus the U(1)$_2$ with $2\varphi$ theory also presumably needs a ``long RG flow" to reach the fixed point. The Dirac theory has manifest $SO(3)$ symmetry and a $U(1)$ gauge field, and thus, is presumably closer to the true fixed point than either bosonic theory. This gives a rationalization for our use of the Dirac theory (rather than the bosonic duals) to make some guesses on the scaling dimensions of physical operators, as we did in the main text.

\section{CDW* to FL transition and tri-critical theory \label{append:FL-CDW*_tri_critical}}   

In the main text we discussed the possibility of a CFL to CDW* transition. In the larger D regime we still expect a FL phase, so there should be a CDW* to FL transition at larger D. In this section, we describe this transition. Essentially the same  problem was studied earlier in Ref. \onlinecite{musser2022theory}. We also point out a theory for the tri-critical point between CFL, CDW* and FL.

\subsection{CDW* to FL transition}

Again we use the slave boson framework, then the CDW* to FL transition corresponds to a superfluid to CDW insulator transition of the slave bosons.  The critical theory in the bosonic sector should have the form

\begin{align}
    S_b&=\int dt d^2x \sum_{i=1,2}|(\partial_\mu -i \alpha_\mu)\varphi_i|^2 +\frac{1}{2\pi} (a + A)  d\alpha-s \varphi^\dagger \varphi + \cdots
    \label{eq:Sb_FL_CDW}
\end{align}
where $\varphi=(\varphi_1,\varphi_2)^T$. The ellipses are quartic terms allowed by the lattice symmetries. $a$ is the U(1) gauge field shared by the slave boson $\Phi$ and the neutral fermion $f$ in the $c=\Phi f$ construction.  The transformation of the two critical bosons $\varphi_1,\varphi_2$ is the same as in eq.~\eqref{eq:Sb_CFL_CDW}.  

The CDW* to FL transition is described by $S_c=S_b+S_f$ with $S_f$ in eq.~\eqref{eq:CFL_FL_Sf}. When $s<0$, the condensation of $\varphi$ leads to a long range CDW order $\vec n=\varphi^\dagger \vec \sigma \varphi$ and Higgses $\alpha$. We are then in the CDW* phase. When $s>0$, $\varphi$ is gapped and integration of $\alpha$ Higgses $a=0$.  Then the fermion $f$ only couples to the probe field $A$ and this is the FL phase. This is a metal-CDW$^*$  insulator transition which is potentially continuous.

\subsection{Tri-critical theory}

We have provided critical theories for phase transitions between each pair of CFL, FL and CDW* phase. Here we provide a tri-critical theory.  The whole action is $S_c=S_b+S_f$ with $S_f$ in eq.~\eqref{eq:CFL_FL_Sf}. The bosonic sector at the tri-critical point is most conveniently described by a SU(2) theory:

\begin{align}
    S_b&=\int dt d^2x\sum_{i=1,2}|(\partial_\mu-i\alpha^s_\mu \tau_s-i \frac{1}{2} a_{\mu} \tau_0 )\Phi_i|^2-s |\Phi|^2 -\frac{1}{4\pi} Tr[ \alpha \wedge d \alpha+\frac{2}{3} i  \alpha \wedge  \alpha \wedge  \alpha]+\frac{1}{8\pi}a d a\notag \\ 
    &~~~- \big(g|\Phi^\dagger \Phi|^2+\lambda_0 \mathbf n \cdot \mathbf n-\lambda (n_4^2+n_5^2))\big)
    \label{eq:SU2_theory_CFL_FL_Sb}
\end{align}

where $\alpha^s_\mu$, $s=0,1,2$ is a SU(2) gauge field.  $\tau_s$ is the pauli matrix in the color space rotating $(\Phi_{a;1},\Phi_{a;2})$ with $a=1,2$ as the flavor index.  $\sigma_a$ is Pauli matrix in the flavor space rotating $\Phi=(\Phi_1,\Phi_2)$.  $|\Phi|^2=\Phi^\dagger \tau_0 \sigma_0 \Phi$.  $\frac{1}{4\pi} \text{Tr}[ \alpha \wedge d\alpha+\frac{2}{3}i \alpha \wedge \alpha \wedge \alpha]$ is the Chern-Simons term for $SU(2)$ gauge field with $\alpha=\sum_s \alpha^s \tau_s$. $a_\mu$ is again the U(1) gauge field also shared with $f$ in $S_f$. There are five bilinear operators $\mathbf n=(n_1,n_2,n_3,n_4,n_5)=( \Phi^\dagger \sigma_1 \Phi, \Phi^\dagger \sigma_2 \Phi, \Phi^\dagger \sigma_3 \Phi,\text{Re} \Phi^T \sigma_2 \tau_2 \Phi, \text{Im} \Phi^T \sigma_2 \tau_2 \Phi)$ in addition to $|\Phi|^2$ in this theory.   
 
 When $s>0$, $\Phi$ is gapped and we have the CFL phase. When $s<0$, $\Phi$ is condensed and we have either the FL or the CDW* phase depending on the condensation pattern controlled by the anisotropy term $\lambda$.  There are two fixed points: $\lambda^*_1>0$ corresponds to CFL-FL transition and $\lambda^*_2<0$ corresponds to CFL-CDW* transition. Both have an emergent SO(3) symmetry.  The tri-critical point corresponds to an unstable fixed point with $\lambda=0$.

\section{Mean-field to realize Chern number changing transition at $\nu=1/4$}
\label{app:mf_14}

Here we explicitly construct the mean-field states for the partons that realize a change of Chern number from $C=1$ to $C=-1$ for $2$ partons simultaneously, at filling $1/4$. This is pertinent to realize the transtion from CFL to FL at physical electron filling $\nu=-3/4$.

The mean field consists of nearest neighbor (NN) hopping between sites $i,j$ with matrix element $u_{ij}$ as
\begin{align}
\label{partonmf}
    u_{i,i+\hat r_1}=\begin{cases}(i)^{i_2+1}e^{i\theta_d} & \textrm{even }i_2\\ (i)^{i_2+1} & \textrm{odd } i_2\end{cases},\nonumber\\
    u_{i,i+\hat r_2}=1,\nonumber\\
    u_{i,i-\hat r_1+\hat r_2}=(i)^{i_2} e^{i\theta}.
\end{align}
where $\hat r_{i}$ are elementary lattice vectors with $60^o$ angle between them and the coordinate for a site $i=(i_1,i_2)$ under such basis. This mean field describes flux $\phi_{\pm}$ for plaquettes in rows with even(odd) $i_2$, respectively, hence flux not invariant under translation along $\hat r_2$. By tuning $\theta$ or $\theta_d$, one could realize $2$ Dirac cones in the mean field and hence transition with Chern number change by $2$ for each of the parton species. One possible choice we find in numerics is $\theta=\frac{29\pi}{80},\theta_d=\frac{\pi}{8}$ and there are $2$ Dirac cones for each parton, realizing a Chern number change from $-1\rightarrow 1$. For $f_{1,2}, \theta_d=0$, $\theta$ could be allowed in a certain range and for convenience we take $\theta=0$ as well.

Take the reciprocal vector associated with $\hat r_{i}$ as $\hat b_i, \hat b_i\cdot \hat r_j=\delta_{ij}$. Translation acts projectively on $f_{1,2}$, i.e.
\begin{align}
    T_1: f_{I,j}\rightarrow f_{I,j+\hat r_1},(I=1,2)\nonumber\\
    T_2: f_{I,j}\rightarrow (i)^{j_1} f_{I,j+\hat r_2}
\end{align}
and $T_1T_2T_1^{-1}T_2^{-1}=i$.

For $f_{3,4}$, we take the mean field for $f_3$ as in eq \eqref{partonmf} with $\theta_d=\pi/8$, and the mean field for $f_4$ is related to that of $f_3$ by a $T_2$ translation, hence they have the same band structure. Since the flux is not translation invariant under $T_2$ with nonzero $\theta_d$, $T_2$ acts by exchanging $f_{3,4}$, to wit
\begin{align}
    T_1: f_{I,j}\rightarrow f_{I,j+\hat r_1},(I=3,4)\nonumber\\
    T_2: f_{3,i}\rightarrow -f_{4,i+\hat r_2},f_{4,i}\rightarrow (-1)^{i_1}f_{3,i+\hat r_2}.
\end{align}
Together, the physical boson $b$ is translation invariant.

\section{Fixing parton and flux densities across the CFL-FL transition}
\label{app:density}

From the action (Lagrangian) in eq \eqref{eq:CFL_FL_whole},\eqref{eq:U1_2psi}, at the CFL-FL critical point, one has $4$ unknown variables: the averages of the flux of $a,\hat a$ and the density of the partons $b,f$. We have, however, $3$ constraints from gauge invariance of $a,\hat a$ and the value of the physical electron density. Thus it might seem that the averages of the flux and parton densities can not be uniquely fixed. In particular, there is no symmetry forbidding a nonzero flux of $a$ as time reversal has been spontaneously broken. Here we derive the detailed constraints and show that for energetic reasons,  the average flux of $a$ will vanish.

We note that the Dirac fermions are defined with a Pauli-Villars heavy fermion band with Chern number $1$, such that the total Chern number is $0,2$ for $m< (>)0$. This means that at the critical point, the density of the Dirac fermion will increase by an additional unit when the flux of $a$ increases by one flux quantum, on top of the contribution from other Chern-Simons term in the action.

Taking derivatives with respect to $a_0,\hat a_0,A_0$ (recall $A_b=A-a$), respectively, one gets the equations of motion, (defining $b=\nabla\times a,\hat b=\nabla\times \hat a$)
\begin{eqnarray}
    \frac{\delta L}{\delta a_0}&=&0:\delta\rho_f=\frac{B+\hat b-b}{2\pi},\nonumber\\
    \frac{\delta L}{\delta \hat a_0}&=&0:\delta\rho_\psi=\frac{B+\hat b-b}{2\pi}-\frac{\hat b}{2\pi}+\frac{\hat b}{2\pi},\nonumber\\
    \frac{\delta L}{\delta A_0}&=&0:B+\hat b-b=0.
\end{eqnarray}
The last term in the middle line is the contribution from the heavy fermion band. The density is measured from the density in the proximate incompressible phases at half-filling for electrons, where $\rho_\psi=0$ and $\rho_f$ corresponds to $1/2$ lattice filling. 

We set $B=0$ (the case of interest) so that  the equations simplify to
\begin{align}
    \hat b=b,\delta\rho_f=\delta\rho_\psi=0.
\end{align}
This fixes the parton densities, though it leaves $b=\hat b$ undetermined.

We resort to energetics to show that $b=\hat b=0$ at the CFL-FL transition.

First the fermions $f$ will have a density response to nonzero flux of $a$ through their Hall conductivity,i.e. $\delta \rho_f=\sigma_{f;xy} b/(2\pi)$, where $\sigma_{f;xy}$ is in general nonzero, and is associated with the Berry curvature of the $f$-fermi surface. As $\delta \rho_f$ is fixed to vanish, this will force $b=0$.

Second, to the lowest order, the free energy density of the boson $\Phi$ and spinon $f$ sectors will change as a function of $b,\hat b$
\begin{align}
    f[b,\hat b]\approx f[b=0,\hat b=0]+\frac{1}{2}\chi_B \hat b^2+\frac{1}{2}\chi_f b^2\nonumber\\
    =f[b=0,\hat b=0] + \frac{1}{2}(\chi_B +\chi_f)b^2,
\end{align}
with $\chi_{B,f}$ the diamagnetic susceptibilities for the parton sectors with positive $\chi_f$, which favors $b=0$ over small but finite $b$.

We can further verify that on two sides of the transition, $b=\hat b=0$. On the fermi liquid side (which corresponds to $m<0$), the equation of motion for $\hat a$ is modified to be $B-b=0$, which fixes $b=0$. This is expected as the $\Phi$ partons are in a superfluid and there will be a Meissner effect for $b$. On the CFL side ($m>0$),the equation of motion for $\hat a$ is modified to be $B+2\hat b-b=0$, which simplifies to $\hat b=0$ given $\hat b-b=0$.
\end{document}